\documentclass[twocolumn]{aastex631}
\graphicspath{{./}{figures/}}
\begin{document}

\newcommand{\vdag}{(v)^\dagger}
\newcommand\aastex{AAS\TeX}
\newcommand\latex{La\TeX}
\newcommand{\WD}{WD~0644+025}
\newcommand{\Msun}{$M_{\odot}$}
\newcommand{\Rsun}{$R_{\odot}$}
\newcommand{\Mearth}{$M_{\oplus}$}
\newcommand{\Rearth}{$R_{\oplus}$}
\newcommand{\Mjup}{M$_{\mathrm{Jup}}$}
\newcommand{\Teff}{$T_{\mathrm{eff}}$}
\newcommand{\Mluna}{M$_{\leftmoon}$}
\newcommand{\logg}{$\log{g}$}
\newcommand{\RWD}{$R_{\text{WD}}$}

\title{The MIRI Excesses around Degenerates (MEAD) Survey II: A Probable Planet detected via IR excess around WD 0644+025}

\author[0009-0008-7425-8609]{Sabrina Poulsen}
\affiliation{Homer L. Dodge Department of Physics and Astronomy, University of Oklahoma, 440 W. Brooks St, Norman, OK 73019, USA}

\author[0000-0002-1783-8817]{John Debes}
\affiliation{AURA for the European Space Agency (ESA), Space Telescope Science Institute, 3700 San Martin Dr, Baltimore, MD 21218, USA}

\author[0009-0002-4970-3930]{Ashley Messier}
\affiliation{Smith College, Northampton MA 01063, USA}

\author[0000-0002-3307-1062]{\'{E}rika Le Bourdais}
\affiliation{Institut Trottier de recherche sur les exoplanètes and Département de Physique, Université de Montréal, 1375 Avenue Thérèse-Lavoie-Roux, Montréal, QC, H2V 0B3, Canada}

\author[0000-0001-9834-7579]{Carl Melis}
\affiliation{Department of Astronomy \& Astrophysics, University of California San Diego, La Jolla, CA 92093-0424, USA}

\author[0000-0002-7698-3002]{Misty Cracraft}
\affiliation{Space Telescope Science Institute, 3700 San Martin Dr, Baltimore, MD 21218, USA}

\author[0009-0003-5977-9581]{Samuel Boucher}
\affiliation{Institut Trottier de recherche sur les exoplanètes and Département de Physique, Université de Montréal, 1375 Avenue Thérèse-Lavoie-Roux, Montréal, QC, H2V 0B3, Canada}

\author[0000-0001-6098-2235]{Mukremin Kilic}
\affiliation{Homer L. Dodge Department of Physics and Astronomy, University of Oklahoma, 440 W. Brooks St, Norman, OK 73019, USA}

\author{Scott Kenyon}
\affiliation{Smithsonian Astrophysical Observatory}

\author[0000-0001-9064-5598]{Mark C. Wyatt}
\affiliation{Institute of Astronomy, University of Cambridge, Madingley Rd, Cambridge CB3 0HA, UK}

\author[0000-0003-3786-3486]{Seth Redfield}
\affiliation{Astronomy Department and Van Vleck Observatory, Wesleyan University, Middletown, CT 06459, USA}

\author{Patrick Dufour}
\affiliation{Institut Trottier de recherche sur les exoplanètes and Département de Physique, Université de Montréal, 1375 Avenue Thérèse-Lavoie-Roux, Montréal, QC, H2V 0B3, Canada}

\author[0000-0003-0475-9375]{Loic Albert}
\affiliation{Institut Trottier de recherche sur les exoplanètes and Département de Physique, Université de Montréal, 1375 Avenue Thérèse-Lavoie-Roux, Montréal, QC, H2V 0B3, Canada}

\author[0000-0001-7106-4683]{Susan E. Mullally}
\affiliation{Space Telescope Science Institute, 3700 San Martin Dr, Baltimore, MD 21218, USA}

\author[0000-0001-8362-4094]{William T. Reach}
\affiliation{Space Science Institute, 4765 Walnut Street, Suite 205, Boulder, CO 80301, USA}

\author[0009-0004-7656-2402]{Fergal Mullally}
\affiliation{Constellation, 1310 Point Street, Baltimore, MD 21231}

\author[0009-0004-9728-3576]{David A. Golimowski}
\affiliation{Space Telescope Science Institute, 3700 San Martin Dr, Baltimore, MD 21218, USA}



\begin{abstract}
The MIRI Excesses Around Degenerates (MEAD) Survey is a Cycle 2 JWST program designed to image nearby white dwarfs with MIRI at 10 and 15~\micron. This survey targeted 56 white dwarfs within 25~pc to search for mid-infrared excesses, flux deficits from collision-induced absorption, and resolved substellar companions. In this paper we present our analysis of \WD, an unusually massive white dwarf (0.95~\Msun) and the MEAD target exhibiting the most significant mid-infrared excess. The observed JWST MIRI photometry shows a 7.3$\sigma$ excess at 15~\micron\ and a 3.6$\sigma$ excess at 10~\micron, which may be associated with either a planetary companion or a circumstellar dust disk. This excess corresponds to a companion mass of 6.8~\Mjup\ (\Teff=261$\pm$9~K) with orbital distance $<$11.8~au, although  substantially lower masses are possible if we consider a closely orbiting insolated companion. No spatially resolved sources are detected within 200~au, with contrast curve analysis excluding planets more massive than 2~\Mjup\ beyond $\sim$12~au. Metal pollution is confirmed in both archival Keck HIRES spectra from 1999 and new observations from 2025, with no evidence suggesting the accretion rate has substantially changed over the decades. We explore possible dust disk morphologies to describe the observed IR excess, and find that traditional debris disks struggle to fit our data. \WD\ thus represents a compelling case study in the growing population of white dwarfs with cold infrared excesses, and highlights JWST’s ability to probe planetary system remnants inaccessible to prior infrared observatories.
\end{abstract}



\section{Introduction} \label{sec:intro}

While the human lifespan is minuscule compared to most astronomical timescales, the abundance of stars observed at varying stages of life allow us to stitch together a picture of their past and future. The evolutionary tracks taken by stars look very different depending on the original mass of the object, yet over 95\% of stars will one day end their life as a white dwarf \citep{fontaine2001}. Despite the importance of studying such a common descendant, our knowledge of the planetary systems hosted by these white dwarfs is poorly constrained. Thousands of exoplanets have been confirmed to date, yet only a handful have been discovered around a white dwarf \citep{vanderburg2020, luhman12, blackman21, zhang2024}. By studying planetary systems at the extremes of their survivability, we gain insight into the architecture and chemistry of planetary building blocks around main sequence stars, as well as insights into the future of our own solar system.

The process by which planets are destroyed or survive their host's evolution is not well understood, however theoretical models do suggest that white dwarfs should host exoplanets. For 1-2 \Msun\ stars, Jupiter mass planets orbiting beyond $\sim$3-4 AU are predicted to survive, while those closer are likely to be engulfed or disrupted by tidal forces during the red giant phase \citep{mustill12}. Massive planets are at greater risk of engulfment than less massive planets orbiting at the same distance, and the radius at which a planet may be engulfed increases with the mass of the star. Any planets that do survive this phase are expected to migrate outward from their original orbit due to stellar mass loss. Radial velocity surveys of giant-branch stars have not shown a significant variation in the frequency of giant planets compared to those around main-sequence stars \citep{wolthoff22}, further suggesting that their survival is likely. It has also been theorized that some WDs may host second-generation planets born of mass lost on the asymptotic giant branch, although these are expected to form mainly in binary systems \citep{hagai10}.

The presence of metals in $\sim$25-50\% of isolated, hydrogen-dominated white dwarfs \citep{zuckerman03,koester14} is another piece of evidence which suggests the existence of surviving planets. As white dwarfs have high surface gravities, any impurities in the atmosphere such as calcium or iron should settle below the observable atmosphere on timescales ranging from thousands of years to a few days \citep{koester09}. Given these timescales are negligible compared to the cooling age of a white dwarf, this implies any impurities we see have recently been accreted onto the star. The origin of these accreted elements was previously attributed to the interstellar medium \citep{dupuis92,dupuis93,hansen03}, yet compelling evidence has accumulated which instead points to asteroids and other rocky planetesimals for the origin of these metals \citep{jura08,jura14}. One widely accepted hypothesis to explain this phenomenon is the presence of giant planets in wide-orbits around the polluted white dwarf, which could perturb the orbits of smaller planetesimals into the roche radius of the white dwarf \citep{alcock86,debes02,jura03}. Observations of distant giant planets orbiting polluted WDs are necessary to test this hypothesis.

Previous searches for these survivors have utilized space telescopes such as Hubble and Spitzer, but were largely unsuccessful in finding planetary candidates. \citet{debes05} searched around seven of the closest DAZs using HST, ruling out companions up to 10-18 \Mjup\ in orbital separations greater than 30 au. \citet{mullally07} used Spitzer to survey 124 white dwarfs, looking for infrared excess and placing limits on companions greater than 10 \Mjup\ less than 30 au from the star \citep[see also][]{kilic09}. \citet{brandner21} used NICMOS to look at seven white dwarfs in the Hyades cluster, constraining companions larger than 7 \Mjup\ beyond 10 au. There have also been successful hunts for these elusive planets- \citet{luhman12}, used Spitzer IRAC to find a cool, low-mass, brown dwarf with common proper motion to a white dwarf. \citet{blackman21} used microlensing techniques to discover a giant planet orbiting a white dwarf. 

Most recently, the unprecedented sensitivity of JWST has enabled the detection of multiple compelling candidates. \citet{mullally24} announced the identification of two resolved candidates using MIRI in Cycle 1, consistent with $\sim$1–7~\Mjup\ at an orbital separation of 0.1–2~au. However, follow up imaging \citep{mullally2025} determined these to be background objects. These early JWST results suggest that broader, systematic surveys can enrich our understanding by uncovering additional companion candidates while also placing statistically meaningful limits on false positive rates. The MIRI Excess Around Degenerates (MEAD) survey (program 3964: PI. S. Poulsen) was developed to carry out a 25~pc volume-limited survey of Gaia-confirmed white dwarfs, using the F1000W and F1500W filters of the MIRI imager. In order to efficiently fill gaps in JWSTs observing schedule, the proposed objects were well distributed across the sky and observations were $<$700~s per star. This shallow imaging design allowed us to observe 56 targets during cycle 2 while still being sensitive enough to detect planets and brown dwarfs in wide orbits. \citet{loic2025} reports the detection of a candidate brown dwarf around the MEAD target 2MASS J09424023$-$4637176. Meanwhile, other Cycle 2 programs have reported similar successes- \citet{limbach2024}, as part of the MEOW survey, identified a mid-infrared excess in the F1800W and F2100W filters around WD 0310–688, consistent with a 1.1–8.5~\Mjup\ companion at an orbital separation of 0.1–2~au.

In this paper we present one of our early MEAD targets with a large infrared excess, \WD. This target was first noted in 1974 under the name G108-26 in a paper searching for cool white dwarfs \citep{hintzen1974}, and was classified as a DA. \cite{zuckerman2003} observed this object on Keck HIRES in 1999 and did not detect a significant amount of metals, calculating limits for the equivalent width of the Ca II K line as $<20~\mathrm{m}\text{\AA}$ and the Ca abundance as [Ca/H] $<$ -11.057. A 2016 paper \citep{barber2016} included \WD\ while cross-correlating several sources of archival photometry for 1265 bright white dwarfs and did not find evidence of an infrared excess based on Spitzer IRAC1 (3.6~\micron) and IRAC2 (4.5~\micron) channels. \cite{bagnulo2021} looked for magnetic white dwarfs in two volume limited samples, and \WD\ was not found to be magnetic.

In this paper, we describe the methods used to obtain our observations, followed by the modeling of the white dwarf's spectral energy distribution. We then present evidence of a candidate unresolved planet, as well as our limits to both unresolved and resolved planet detection. Finally, we present Keck HIRES spectra supporting the reclassification of \WD\ as metal polluted, and consider if a dusty disk could be the source of the IR-excess. The paper concludes with a summary of our primary findings and a discussion of relevant caveats and limitations that should be considered when interpreting these results.

\section{Observations} \label{sec:sec2}
\subsection{JWST MIRI} \label{sec:miriobs}
\begin{figure}[ht]
\begin{center}
    %\plotone{ColorImage2.png}
    \includegraphics[width=\columnwidth]{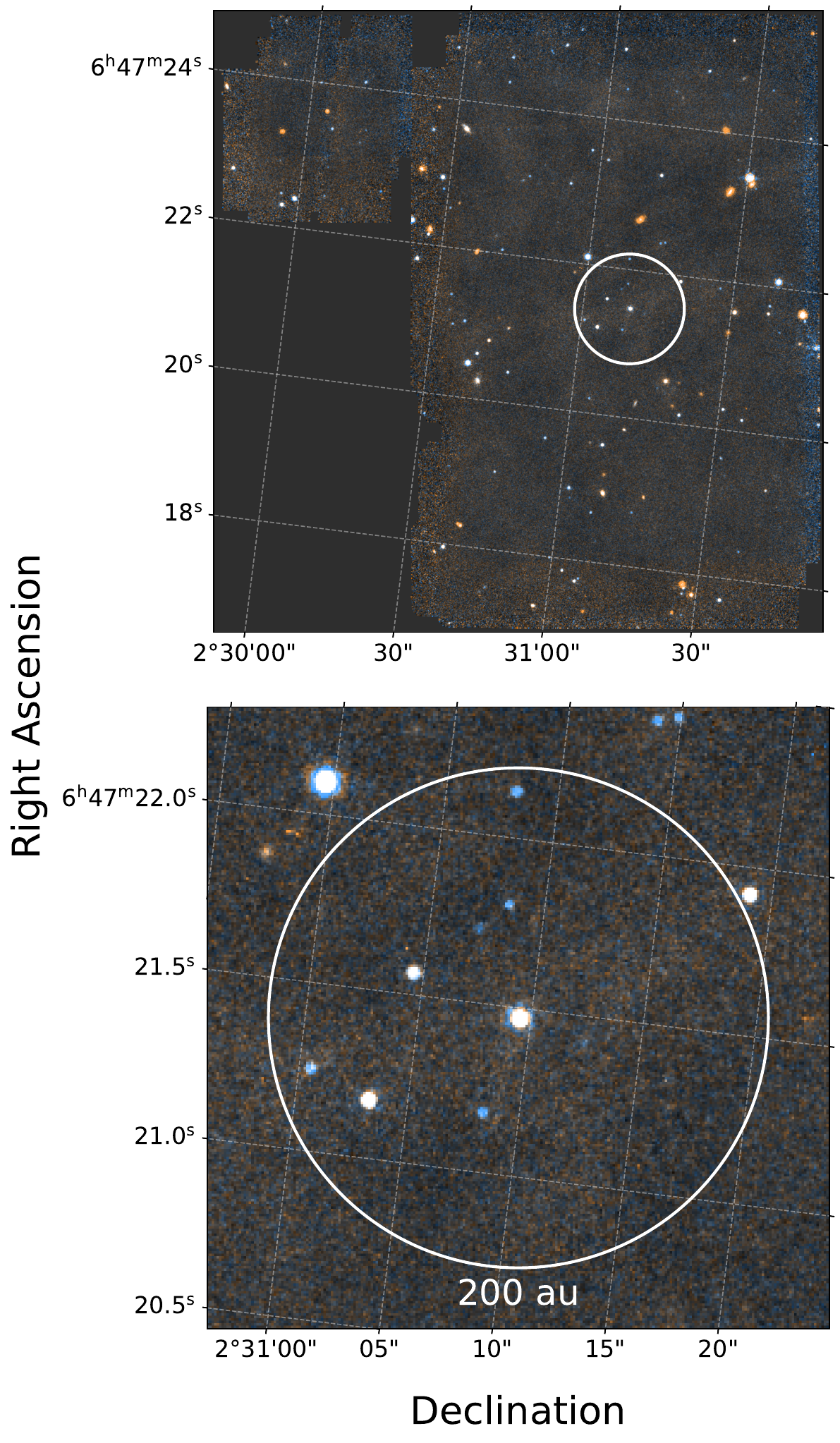} 
    \caption{A squared-stretch false color image of \WD, constructed from F1000W (blue), F1500W (red), and a green channel formed from the average of the two bands. The upper panel displays the full MIRI field, and the lower panel shows a magnified view centered on the target. The white circle marks a projected radius of 200~au. No resolved point-source candidates are identified within this region.}
    \label{fig:colorImage}
\end{center}
\end{figure}

JWST observed \WD\ on 2024 March 25 in two filters, F1000W and F1500W (data can be found in MAST: \dataset[https://doi.org/10.17909/m7zd-8c20]{https://doi.org/10.17909/m7zd-8c20}). Each exposure was dithered using the 4 point Cycling dither pattern with the FASTR1 readout, with a total exposure time of 222~s in F1000W and 444~s in F1500W. The data were processed with build 11.1 \citep{bushouse2023} of the JWST calibration pipeline using the standard stages calwebb\_detector1, calwebb\_image2, and calwebb\_image3. In calwebb\_detector1 and calwebb\_image2, default parameters were used except for the jump step, where the rejection threshold was set to 5.0. After calwebb\_image2, the individual *cal.fits images were stacked to create a median sky image, which was subtracted from each frame to remove both background emission and residual detector effects. In calwebb\_image3, a ‘square’ resample kernel was used and the weight\_type was set to ‘exptime’. The outlier detection step employed ‘scale’ values of 1.0 and 0.8—double the default values—to better identify and mask cosmic rays and related pixel artifacts. The final output consisted of a single combined and resampled image per filter (i2d.fits), with sky levels near zero. These images form the basis for all subsequent analysis. Fig. \ref{fig:colorImage} shows a color composite image of the 10 and 15~\micron\ exposures.

\subsection{Keck HIRES} \label{sec:hiresobs}

Modern-epoch observations of \WD\ were obtained with the Keck~I telescope and HIRES at Maunakea Observatory \citep{vogt1994}. One viable 50~minute exposure was obtained on UT 23 November 2024 and two 40~minute exposures were obtained on UT 18 February 2025. HIRES data were taken with the C5 decker (1.148" slit width; R$\sim$40,000) and used setups that covered 3115-4000~\AA, 4055-5000~\AA, and 5050-5950~\AA\ across the three detectors. Data are reduced using the {\sf MAKEE} software package that outputs heliocentric velocity-corrected spectra shifted to vacuum wavelengths. Reduced one-dimensional spectra produced by {\sf MAKEE} are normalized by dividing by high-order polynomials fit to the continuum. The resulting coadded data, produced by weighted averaging of all exposures and order overlaps, are hereafter referred to as the 2025 spectra.

\WD\ was previously observed with HIRES as described in \cite{zuckerman2003}. This data, referred to as the 1999 spectra, is compared in detail to the 2025 spectra in Section 4.3.

\section{White Dwarf Modeling and Photometry}

To assess whether \WD\ is orbited by unresolved planets or a dust disk, we first construct an accurate model of its spectral energy distribution (SED) using available photometric data. By subtracting the modeled white dwarf flux at infrared wavelengths, we can isolate any potential excess indicative of a cooler companion or circumstellar dust. This analysis begins with the collection and fitting of observed photometry spanning the visible, near-infrared, and mid-infrared bands. Photometric data were compiled from Pan-STARRS \citep{chambers2016}, Gaia DR3 \citep{Gaia2023}, the 2MASS Point Source Catalog \citep{cutri2003}, WISE W1 and W2 bands through the CATWISE catalog (non-detection in W3 and W4) \citep{eisenhardt2020}, and Spitzer IRAC 1 and 2 through the GLIMPSE-360 catalog \citep{winston2020}.

% SDSS $g'$/$r'$/$i'$/$z'$ \citep{kollmeier2019},

\begin{deluxetable*}{lccccrr}
\tablehead{
\colhead{Filter} &
\colhead{Central Wavelength} &
\multicolumn{1}{c}{$F_{\mathrm{Modeled}}$} &
\multicolumn{1}{c}{$F_{\mathrm{Observed}}$} &
\multicolumn{1}{c}{$F_{\mathrm{Excess}}$} &
\multicolumn{1}{c}{$\sigma_\nu$} &
\multicolumn{1}{c}{$F_{\mathrm{Excess}}/\sigma_\nu$} \\
\colhead{} &
\colhead{$\mu$m} &
\multicolumn{1}{c}{mJy} &
\multicolumn{1}{c}{mJy} &
\multicolumn{1}{c}{mJy} &
\multicolumn{1}{c}{mJy} &
\multicolumn{1}{c}{}
}
\decimals
\tablecaption{Photometry of \WD}
\label{tab:photparams}

\startdata
Johnson B                  & 0.4335 & 1.5297 & 1.5271 & -0.0026 & 0.209 & -0.012 \\
Johnson V                  & 0.5454 & 1.9486 & 2.0607 & 0.1121 & 0.107 & 1.05 \\
Pan-STARRS g               & 0.4832 & 1.7412 & 1.7865 & 0.0453 & 0.0179 & 2.53 \\
Pan-STARRS r               & 0.6188 & 2.0392 & 2.0324 & -0.0068 & 0.0203 & -0.35 \\
Pan-STARRS i               & 0.7525 & 2.1030 & 2.0701 & -0.0329 & 0.0207 & -1.59 \\
Pan-STARRS z               & 0.8669 & 2.0406 & 1.9953 & 0.0453 & 0.0200 & 2.27 \\
Pan-STARRS y               & 0.9622 & 1.9567 & 1.9770 & 0.0203 & 0.0198 & 1.03 \\
Gaia DR3 G$_{\mathrm{BP}}$ & 0.5129 & 1.7678 & 1.7435 & -0.0243 & 0.0523 & -0.46 \\
Gaia DR3 G                 & 0.6425 & 2.0082 & 1.9981 & -0.0101 & 0.0599 & -0.17 \\
Gaia DR3 G$_{\mathrm{RP}}$ & 0.7799 & 2.0969 & 2.0539 & -0.0430 & 0.0616 & -0.70 \\
2MASS J                    & 1.2378 & 1.6936 & 1.8156 & 0.1220 & 0.0817 & 1.50 \\
2MASS H                    & 1.6476 & 1.3198 & 1.2848 & -0.0350 & 0.0887 & -0.39 \\
2MASS K$_s$                & 2.1620 & 0.8688 & 0.9864 & 0.1176 & 0.102 & 1.15 \\
CATWISE W1                 & 3.3792 & 0.4134 & 0.4722 & 0.0588 & 0.0142 & 4.14 \\
CATWISE W2                 & 4.6293 & 0.2297 & 0.2257 & -0.0040 & 0.0072 & -0.56 \\
Spitzer IRAC 1             & 3.5573 & 0.3710 & 0.3960 & 0.0250 & 0.0119 & 2.10 \\
Spitzer IRAC 2             & 4.5050 & 0.2403 & 0.2460 & 0.0057 & 0.0074 & 0.77 \\
MIRI F1000W                & 9.9623 & 0.0513 & 0.0556 & 0.0043 & 0.0012 & 3.58 \\
MIRI F1500W                & 15.1003 & 0.0228 & 0.0337 & 0.0109 & 0.0015 & 7.27 \\
\enddata

\end{deluxetable*}

We use the photometric technique as detailed in \citet{bergeron2019}, where we use the available optical and infrared photometry along with the Gaia DR3 parallaxes to constrain the effective temperature and the solid angle. Since the distance is known, we constrain the radius of the star directly, and use white dwarf evolutionary models to calculate the mass. We convert the observed magnitudes into average fluxes, and compare with the synthetic fluxes calculated from model atmospheres. We minimize the $\chi^2$ difference between the observed and model fluxes over all band passes using the nonlinear least-squares method of Levenberg-Marquardt \citep{press1986} to obtain the best fitting parameters. The uncertainties of each fitted parameter are obtained directly from the covariance matrix of the fitting algorithm, while the uncertainties for all other quantities derived from these parameters are calculated by propagating in quadrature the appropriate measurement errors.

In addition to the photometric fit, we performed a spectroscopic analysis using the Keck HIRES data. Starting from the effective temperature and surface gravity obtained from the photometric fit, we determine the Ca abundance following the method in \cite{dufour2012} on all available spectra. We explored a grid of pure H models spanning $\log{\rm Ca/H}=-12$ to $-3$.  The best-fitting parameters are obtained by minimizing the $\chi^2$ difference between the observed and synthetic spectra. The final abundance was obtained by averaging the abundances calculated on each individual line.

\begin{figure*}
\begin{center}
\plotone{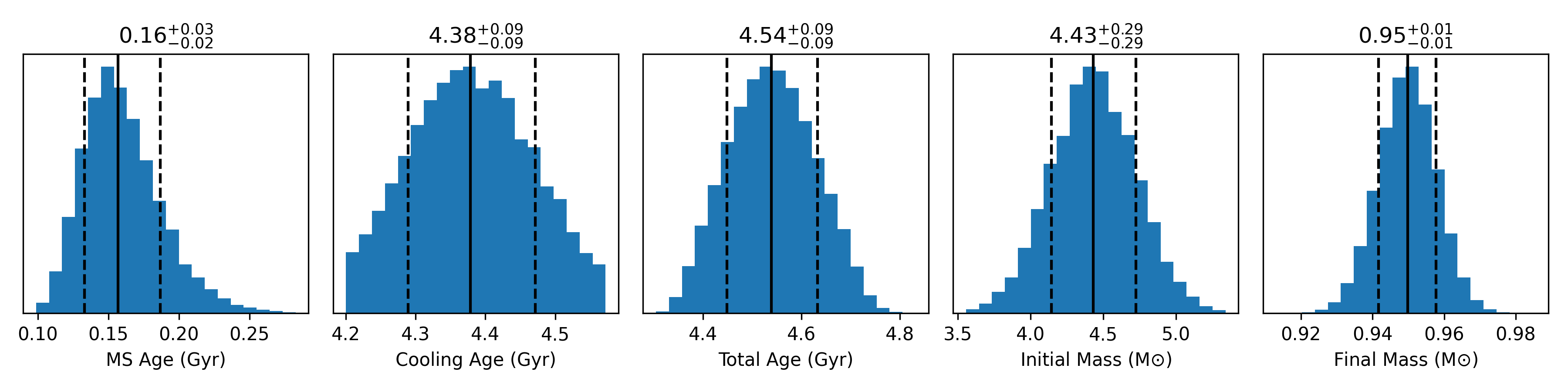}
\caption{Probability distributions for the age, initial mass, and final mass of \WD\ derived using wdwarfdate \citep{kiman22}. The values \Teff=6943$\pm$68~K and log~g=8.555$\pm$0.013 were used for input values. The DA cooling models \citep{bedard20}, MIST based IFMR \citep{cummings18}, and [Fe/H]=0.0, v/vcrit=0.0 \citep{choi16, dotter16} were chosen for the model WD, model IFMR, and isochrone of the progenitor star respectively.}
    \label{fig:wddate}
\end{center}
\end{figure*}

Using the photometric technique, we find \WD\ has an effective temperature of 6943$\pm$68~K and a log~g of 8.555$\pm$0.013. Using the spectroscopic technique, we find an effective temperature of 6870$\pm$82.66~K and a log~g of 8.62$\pm$0.039. These values are broadly consistent with previous analyses, such as \citet{obrien2024}, who used Balmer line profile fitting to derive \Teff\ = 7109$\pm$50~K and log~g = 8.59$\pm$0.01, and \citet{bedard2017}, who also employed Balmer line fitting and found \Teff\ = 7085$\pm$106~K and log~g = 8.52$\pm$0.07. \citet{blouin2019} used a combination of photometric and spectroscopic fitting with updated atmosphere models for cool white dwarfs, reporting \Teff\ = 6995$\pm$60~K and log~g = 8.58$\pm$0.01. As the photometric fit is constrained over a broader wavelength range and provides a more reliable representation of the photosphere at longer wavelengths, we adopt the photometric parameters as our final atmospheric solution. The mass and total age are then inferred from the DA \cite{bedard20} cooling model and the WD mass-radius relationship using \Teff\ = 6943$\pm$68~K and log~g = 8.555$\pm$0.013 (Fig \ref{fig:wddate}). This implies a mass of 0.95~\Msun\ and a total age of 4.54~Gyr.

To measure the photometry of \WD\ in each MIRI filter, we performed aperture photometry using the recommended aperture and background radii from the CRDS calibration reference data, as specified in the MIRI aperture correction file \citep{jdox16}. The chosen photometric apertures enclose approximately 80\% of the target’s total flux. We add a 2\% systematic uncertainty in quadrature to account for limitations in our understanding of white dwarf atmospheres in the mid-infrared. This estimate is based on comparisons between MIRI photometry and well-characterized white dwarfs. A more detailed description on measuring the photometry can be found in \cite{debes2025}.

To predict the expected flux of the white dwarf for each filter, we interpolated the existing models to the reference wavelengths of both MIRI filters. Using the best fit \Teff\ and log~g for \WD, we generated a high resolution photospheric model extending to 30\micron. We then scaled the photospheric model by the best fit radius and distance. Finally, we used the publicly available MIRI filter transmission profiles from the Spanish Virtual Observatory Filter Service \citep{rodrigo2024} to convolve the models with each filter profile to create an accurate prediction of the MIRI flux in each band \citep{holberg06}. Table \ref{tab:photparams} gives the modeled and observed photometry for archival and newly obtained observations. We find that there is a slight 3.6$\sigma$ excess in F1000W, and a large 7.3$\sigma$ excess in F1500W. The implications of this excess is discussed in the Unresolved Detection and Limits subsection of the results (\S\ref{sec:unresolveddetection}).

\section{Results} \label{sec:results}

 In this section, we present evidence for a significant IR excess that may be associated with either a planetary companion or a circumstellar dust disk. We discuss our search for resolved planetary companions as well as our analysis of background objects in the MIRI field. We also present newly taken Keck HIRES data showing \WD\ as metal polluted, as well as implied accretion rates. Finally, we compare our IR excess with several dust disk models and evaluate their consistency with current observational constraints.
 
\subsection{Infrared Excess}
\label{sec:unresolveddetection}

\begin{figure}[h!]
\begin{center}
    %\plotone{sed_fig_wd2149.eps}
    \includegraphics[width=\columnwidth]{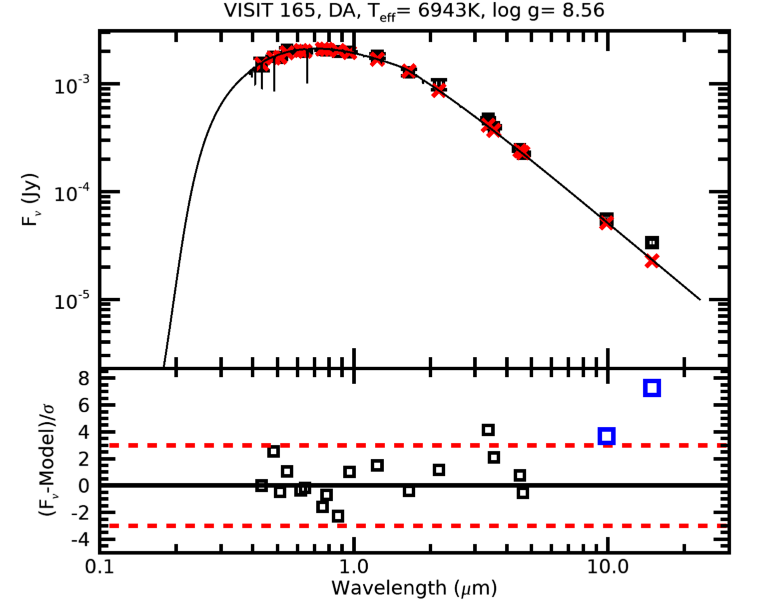}
    \caption{Top: The measured spectral energy distribution of \WD\ (black squares) compared to a DA cooling model with \Teff=6943~K and \logg=8.56 (red squares). Bottom: The residuals between modeled and observed photometry in units of sigma. The observed JWST MIRI photometry is shown with the blue squares and archival photometry is shown in black squares. The red dashed line denotes a 3$\sigma$ deviation from the model. The observed JWST MIRI photometry exceeds the expected photospheric flux by 3.6$\sigma$ in F1000W and 7.3$\sigma$ in F1500W.} 
    \label{fig:sed}
\end{center}
\end{figure}

After plotting the model spectrum on top of our observed flux (Fig. \ref{fig:sed}), we see that the observed JWST MIRI photometry shows a 7.3$\sigma$ excess in the F1500W filter, and a 3.6$\sigma$ excess in the F1000W filter. We also note an excess above 3$\sigma$ in the CATWISE 1 photometry, however given the higher spatial resolution of Spitzer IRAC 1 we consider the later photometry \citep{winston2020} more believable. As the excess is unresolved, the orbital separation is constrained to be $\leq$12~au, while no meaningful limit can be placed on the inner boundary. We therefore explore a continuous range of planet masses and separations consistent with the observed mid-infrared excess, allowing for contributions from both internal heat and white dwarf irradiation.

We construct an isochrone using cloudless Sonora-Bobcat evolutionary models \citep{SonoraBob} for companions more massive than 2~\Mjup, and cloudless Helios (Bex) models \citep{linder19} for those below this threshold. We then identified the corresponding \Teff\ and \logg\ values associated with various substellar companion masses from the Sonora-Bobcat grid. Using these parameters, we interpolated the JWST predicted fluxes from the model grids at the age of \WD, 4.54 Gyr. These model fluxes were then converted to apparent fluxes by accounting for the distance to \WD, 18.09~pc. We can then compare the measured F1500W excess to the modeled grid to infer the characteristic temperature of a companion consistent with the observed mid-infrared emission, finding that it is best matched by a model with $T_{\mathrm{eff,pl}}$=261$\pm$9~K. For this temperature, the corresponding model prediction in F1000W is 1.80~$\mu$Jy, which deviates from the observed flux by $\sim$2.1$\sigma$.

%We consider the smallest mass this observation should have been sensitive to by interpolating our isochrone on a hypothetical 3$\sigma$ excess in F1500W. This corresponds to F1500W flux of 0.0273~mJy, and a planet mass of 4.65~\Mjup.

%\textbf{For the second case, in which heating from the white dwarf dominates the companion’s temperature, we quantify the effect by identifying the combinations of planetary temperature and radius whose modeled flux agrees with the observed excess to within 1$\sigma$.} For a given temperature and wavelength, equation \ref{eq:model_flux} gives the modeled flux in terms of planet radius (R).

We treat the planet’s effective temperature as the combination of an intrinsic component set by the companion mass $m_{\mathrm{pl}}$ and an external component due to insolation at orbital separation $a$, such that

\begin{equation}
T_{\mathrm{eff,pl}} = \left[T_{\mathrm{int}}\!\left(m_{\mathrm{pl}}\right)^{4} + T_{\mathrm{ext}}(a)^{4}\right]^{1/4}
\label{eq:Teff_pl}
\end{equation}

%\begin{equation}
%F_{\text{model}} = B(T, \lambda) \cdot \Omega = B(T, \lambda) \cdot \pi \frac{R^2}{D^2}
%\label{eq:model_flux}
%\end{equation}

%A range of radii can then be tested to see which values satisfy the condition $\left|F_{\text{model}} - F_{\text{excess}}\right| \leq F_{\text{err}}$. We find that temperatures as low as 191~K fit our observed flux excess for $R=1.7-1.75~R_{\text{Jup}}$, while temperatures as high as 277~K are valid for $R=0.69~R_{\text{Jup}}$. For planetary radii below 1.1~$R_{\text{Jup}}$, the approximate relationship 0.39~\Mjup$\times(R_{pl}/12.1~R_\oplus)^{1.8}$ \citep{bashi17} estimates a mass of M(R=0.69~$R_{\text{Jup}}$) = 0.18~\Mjup. Finally, equation \ref{eq:orbital_dist} allows us to recover an orbital distance, assuming $\alpha$=0.3. For $T_{\text{eq}}$ = 191~K and 277~K, $a$=0.033~au (7.1~\Rsun) and 0.016~au (3.4~\Rsun) respectively. 

The external temperature due to white dwarf irradiation is defined as

\begin{equation}
T_{\text{ext}}(a) = T_{\text{eff,WD}} \left(1 - \alpha\right)^{1/4} \sqrt{\frac{R_{\text{WD}}}{2a}}
\label{eq:orbital_dist}
\end{equation}

\begin{figure}[]
\begin{center}
    %\plotone{sed_fig_wd2149.eps}
    \includegraphics[width=\columnwidth]{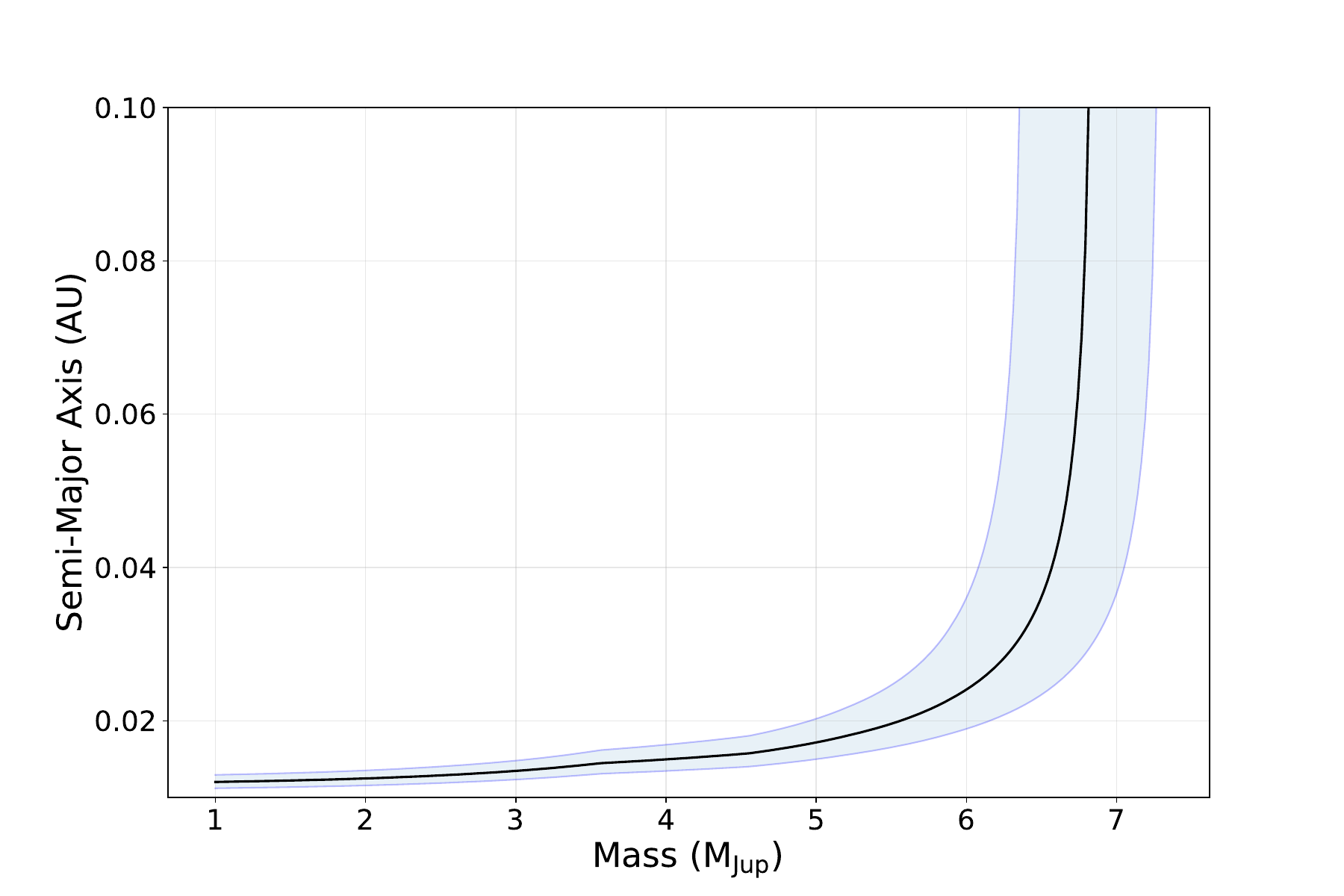}
    \caption{Orbital distance as a function of companion mass consistent with the observed mid-infrared excess under the combined intrinsic and insolation heating model (Eq.~\ref{eq:Teff_pl} and \ref{eq:orbital_dist}). The black curve shows the best-fit solution corresponding to $T_{\mathrm{eff,pl}}$=261~K, while the blue curves and shaded region indicate the range spanned by the 1~$\sigma$ bounds ($T_{\mathrm{eff,pl}}$=252, 270~K).} 
    \label{fig:smaVSmass}
\end{center}
\end{figure}

To explore the range of companion masses and orbital separations consistent with the observed mid-infrared excess, we evaluate Equations \ref{eq:Teff_pl} and \ref{eq:orbital_dist} across a finely sampled grid of internal temperatures. For each $T_{\mathrm{int}}$, we solve for the orbital separation required to reproduce the inferred effective temperature, $T_{\mathrm{eff,pl}}$=261$\pm$9~K. The internal temperature can then be converted into a planetary mass by interpolating on our combined isochrone. We note that this analysis is limited by the availability of evolutionary models at low masses and temperatures, with the minimum mass in the model grid being 1~\Mjup\ (corresponding to $T_{\mathrm{int}}\sim124$~K). This procedure yields a continuous semi-major axis vs mass relation (Figure \ref{fig:smaVSmass}), with a 1~\Mjup\ companion corresponding to a separation of $a$=0.012~au, a 3~\Mjup\ companion corresponding to $a$=0.0135~au, and a 6~\Mjup\ companion corresponding to $a$=0.024~au. External heating becomes negligible by $a\sim0.10$~au as $T_{\mathrm{int}}$ approaches $T_{\mathrm{eff,pl}}$, and the inferred mass remains near the asymptotic value of 6.8~\Mjup\ out to the unresolved limit of 12~au. 

%\textbf{These two scenarios provide an upper and lower bounds on the planet’s mass, ranging from 0.18 to 6.8 $M_{\text{Jup}}$. The precision of these estimates is limited by uncertainties in parameters such as the planetary albedo and by the approximate nature of the assumed radius–mass relation. Even so, the derived range represents a reasonable and physically consistent estimate of the companion’s likely mass.}

\subsection{Resolved Companion Detection Limits} 
\label{subsec:resolvedDetection}

A systematic analysis of potential false positives was conducted across all MEAD fields by examining the photometric colors and FWHM morphology of background sources within 200~au of each white dwarf. Objects were flagged as planet or brown dwarf candidates if they exhibited red colors consistent with substellar temperatures and FWHM profiles suggesting point-like geometry. \WD\ was not identified as containing any such sources. Analyzing the entire field for all MEAD targets, the contamination rate of red, point-source objects is approximately 0.002~$\text{arcsec}^{-2}$. For \WD, this corresponds to a $\sim$76\% probability that at least one background object lies within a 200~au radius. We adopt 200~au as our standard search region for the entire MEAD sample to minimize false positives, although white dwarfs have been found with companions at separations as high as 2500~au \citep{luhman12}. 

To complement the population-level screening, a grid search was also performed for this white dwarf. For each filter, we defined an annular region extending from four times the filter-specific FWHM to a maximum radius of 200~au. Within this region, we placed circular apertures of radius $0.67 \times \mathrm{FWHM}$ centered on each pixel and measured the enclosed flux. The distribution of aperture sums was modeled with a Gaussian to estimate the typical background level, which was then subtracted from each aperture sum to isolate excess signals. A search for apertures which deviated more than 5$\sigma$ from the mode was conducted, identifying only known background sources.

While no resolved candidates were identified, we can estimate the least massive planet our observation should have been sensitive to by constructing a contrast curve. This compares the contrast between a circular aperture located at the center of the white dwarf against multiple equally sized apertures located a fixed distance away from the white dwarf. A more detailed description of this procedure can be found in \cite{poulsen24}, with the exception that a new method was used to remove background sources. The prior method involved calculating the median absolute deviation for each ring of apertures such that apertures with deviations greater than 5$\sigma$ were rejected and did not contribute to the contrast curve. This method consistently rejected apertures which strongly overlapped with background sources, however apertures which only partially overlapped were frequently not rejected. By retaining apertures with even minimal overlap, the true variation in the background is overestimated.

In contrast, the rejection method presented in this work functions by directly removing outlier pixels. An annulus centered on the white dwarf is defined with $r_{\mathrm{inner}} = 4 \times r_{\mathrm{aperture}}$ and $r_{\mathrm{outer}} \sim\ 200$ au, corresponding to the maximum distance probed by the contrast curve. We then plot the value of every pixel within this annulus and fit the distribution to a Gaussian. Any pixels with values outside the range $\pm 3\sigma$ are converted to NaN. If any aperture includes a NaN value it is automatically rejected. Overall, this approach yields results that closely align with the original method while offering modest improvements near the edges of background sources. The distribution of measured photometry in the remaining apertures is roughly Gaussian, consistent with a well-behaved background.

\begin{figure}[]
\begin{center}
    \includegraphics[width=\columnwidth]{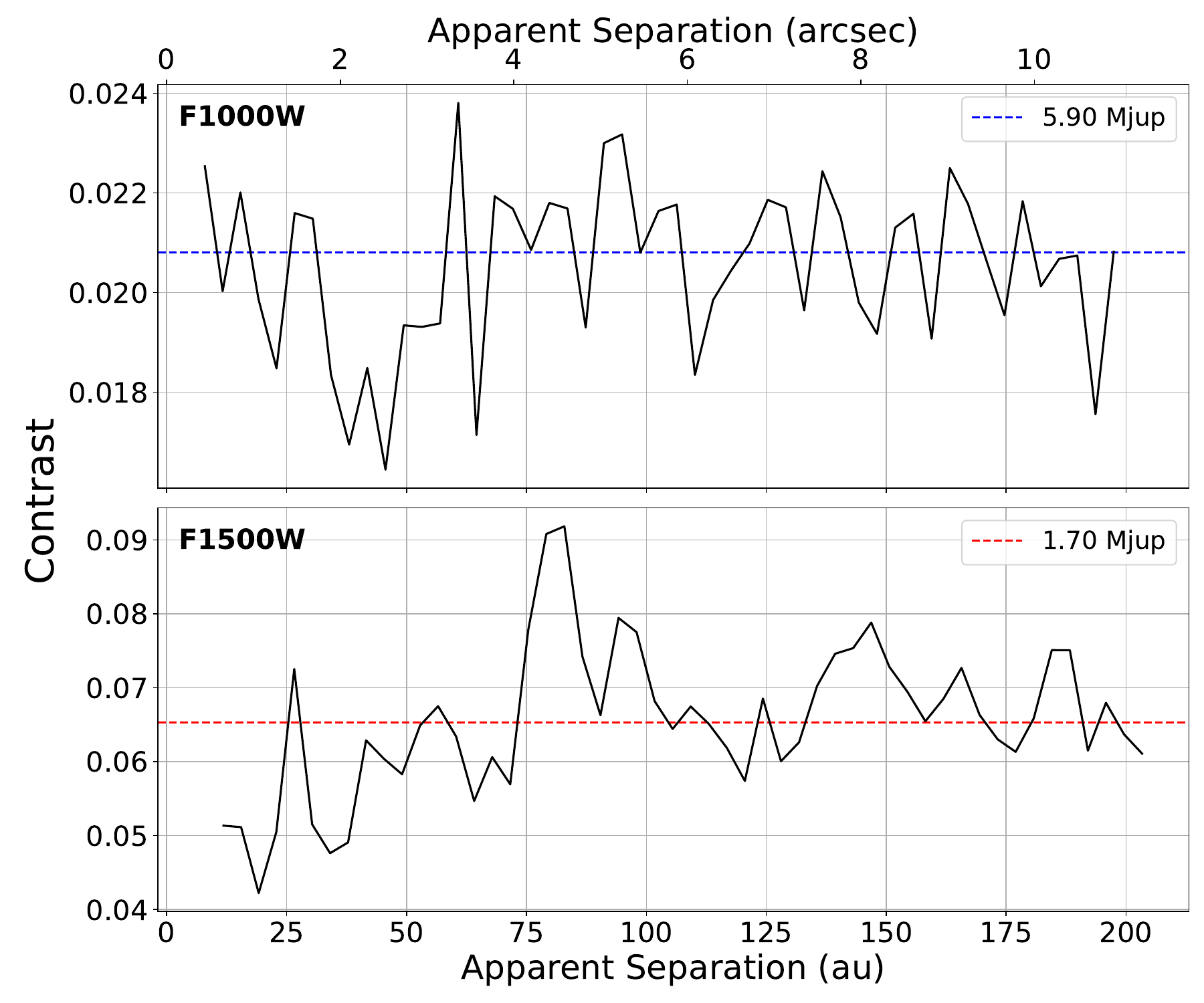}
    \caption{This contrast curve reports the flux ratio of our central aperture relative to the standard deviation of the background at varying radial distances for our two filters, F1000W (top) and F1500W (bottom). The dashed lines represent median contrast values of 0.0208 (F1000W) and 0.0653 (F1500W). These contrasts are reported as a limiting planet mass calculated by interpolating on the 4.54~Gyr Bex model. The inner most working angle is 7.95~au in F1000W and 11.83~au in F1500W.}
    \label{fig:ccCombined}
\end{center}
\end{figure}

To obtain a meaningful contrast near the white dwarf, reference PSFs were constructed using LAWD 26 (the brightest and highest signal-to-noise object in the MEAD survey) and scaled before being subtracted from \WD. This PSF subtraction dramatically improved contrast in the F1000W filter interior to $\sim$0.85\arcsec (15.38 au). In contrast, the F1500W filter showed minimal improvement, as the white dwarf is faint enough at this wavelength that the first Airy ring was barely detected in the sky-subtracted image. For F1000W, the median contrast remains nearly constant across all regions. However, in F1500W, the contrast initially increases with distance from the white dwarf, likely a result of structured background emission at longer wavelengths. The final 5$\sigma$ PSF-subtracted contrast curves for both filters are presented in Figure \ref{fig:ccCombined}. To convert the median contrast into planetary mass estimates, we again use the isochrone described in \S\ref{sec:unresolveddetection}. In F1500W, 86\% of radial distances are sensitive to 2~\Mjup\ planets or better.

%\begin{deluxetable*}{rccc}
%\label{tab:wdcparams}
%\tablehead{
%\colhead{Filter} & \colhead{Inner Working Angle} & \colhead{Median Contrast} & \colhead{Limiting Mass}\\ [-0.2cm]
%\colhead{} & \colhead{au} & \colhead{} & \colhead{\Mjup} \\[-0.5cm]}
%\startdata
%F1000W  & 7.95 & 0.02080 & 5.90 \\
%F1500W & 11.83 & 0.06530 & 1.70 \\
%\enddata
%\caption{Median contrast values for each filter at near (F1000W: 8-33~au, F1500W: 12-37~au), mid (F1000W: 36-66~au, F1500W: 40-68~au) and far (F1000W: 68-99~au, F1500W: 71-102~au) radii. The median values are calculated based on the PSF subtracted contrast curve.} 
%\end{deluxetable*}

\subsection{Metal Abundance and Accretion Rate} \label{subsec:metalAbundance}

\begin{figure}[]
    \centering
    \includegraphics[width=\columnwidth]{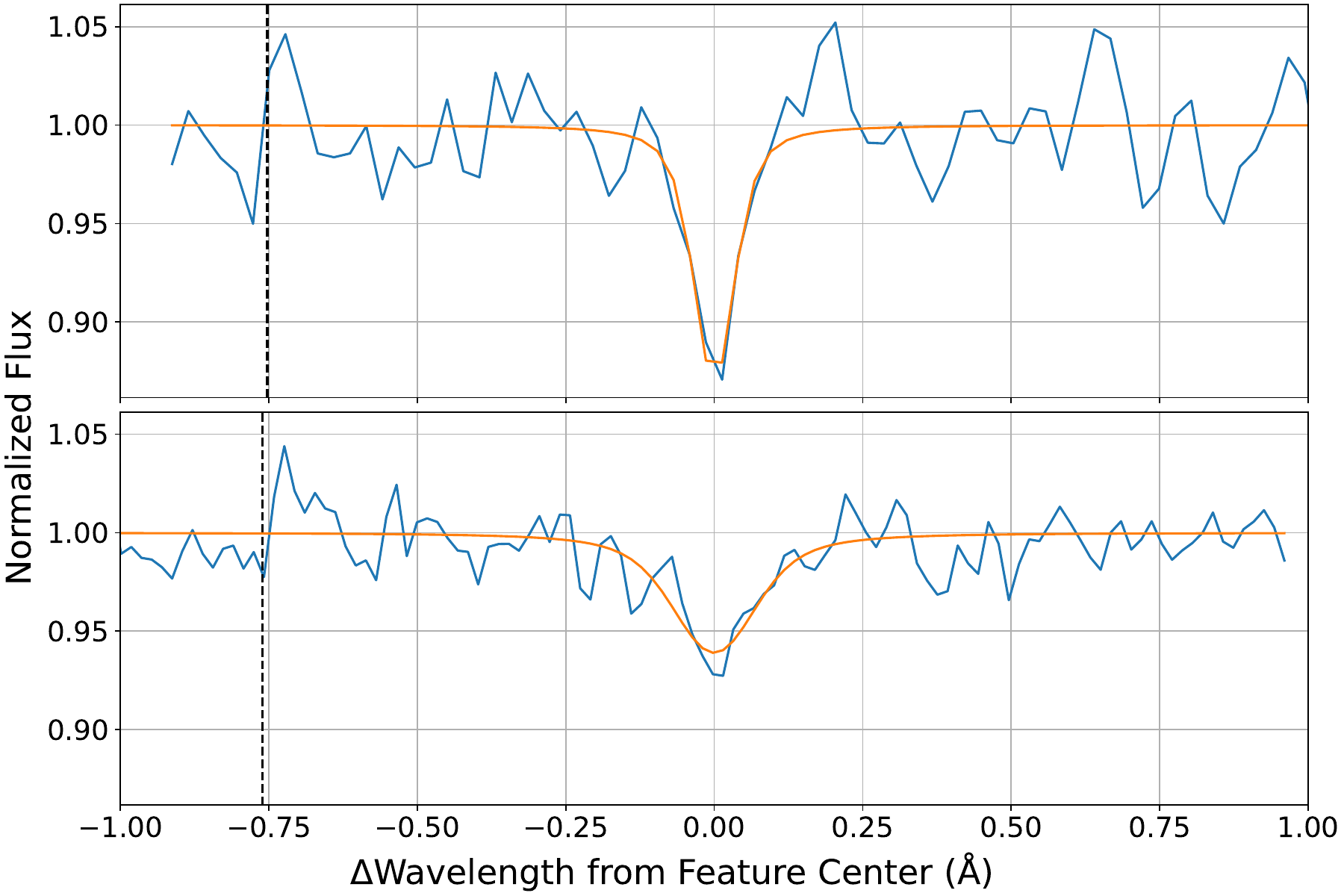}
    \caption{Comparison of the Ca~II absorption feature in the Keck HIRES spectra of WD~0644+025 from 1999 (top) and 2025 (bottom). For analysis, both spectra are smoothed using a 1D boxcar kernel with a width of 3 pixels and normalized to the local continuum approximately 4~\AA\ on either side of the line. The unsmoothed, continuum-normalized spectrum is shown as the blue curve, and the best-fit Voigt profile is shown as the orange curve. Both epochs are calibrated using an air wavelength solution, and the dashed vertical line indicates the Ca~II rest wavelength at 3933.663~\AA.}
    \label{fig:ca_feature_combined}
\end{figure}

We compare Keck HIRES spectra of \WD\ obtained in 2024/2025 to archival observations from 1999. Both spectra were smoothed using a 1D boxcar kernel with a width of 3 pixels and normalized to the continuum approximately 4~\AA\ on either side of the Ca II feature. Figure \ref{fig:ca_feature_combined} displays the unsmoothed spectra, best-fit models, and the Ca II rest wavelengths in an air wavelength solution. In addition to Ca II, we searched for other common metallic species but found no detectable features. Equivalent widths and line parameters are measured using a 1000-iteration bootstrap procedure. Each iteration resamples the spectrum with replacement to create a synthetic dataset, which is then fit with a Voigt profile using LMFIT \citep{lmfit}. The median and 1~$\sigma$ confidence intervals of the resulting parameter distributions are adopted as the reported values and uncertainties. We measure radial velocities of $57.30^{+0.15}_{-0.35}$~km/s and $57.87^{+0.45}_{-0.53}$~km/s for the 1999 and 2025 data, respectively, and find no evidence of significant change between epochs. The equivalent widths of the Ca II feature are also consistent, measured at $13.7^{+0.6}_{-1.5}$~m\AA\ in 1999 and $15.2^{+1.2}_{-1.1}$~m\AA\ in 2025, providing no evidence for a change in the Ca accretion rate. 

\begin{figure*}[t]
    \centering
    \includegraphics[width=\textwidth]{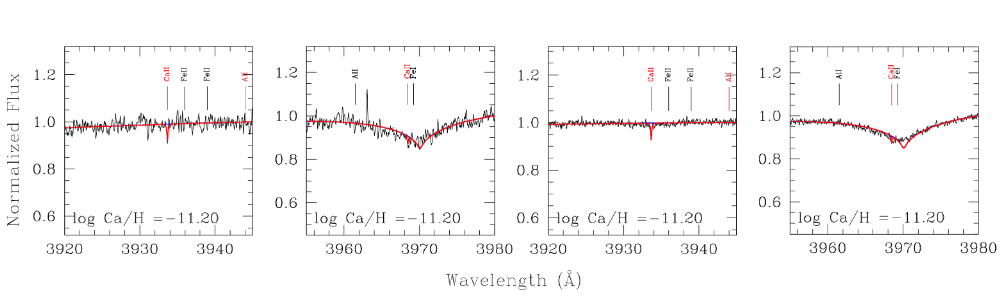}
    \caption{Normalized HIRES spectrum (black) with our best-fit model (red). The two left panels show the 1999 dataset, and the two right panels show the 2025 dataset.}
    \label{fig:Ca_abundance}
\end{figure*}

The Ca abundance was determined following \cite{dufour2012} on all available spectra (Fig \ref{fig:Ca_abundance}). The final abundance was obtained by averaging the abundances calculated on each individual line. This Ca abundance was determined to be [Ca/H]$\mathrm{=-}11.20\pm0.1$ dex. \cite{zuckerman03} did not identify a Ca line when publishing the 1999 spectra due to the noise level of the data, but did include a limit for the equivalent width as $<20~\mathrm{m}\text{\AA}$ and limited the Ca abundance to [Ca/H] $<$ -11.057. We note that our equivalent width and Ca abundance are consistent with these limits.

The Ca abundance can be used to find an implied Ca accretion rate $\dot{M}(\mathrm{Ca})$, which is calculated using the following equation, derived in \cite{koester09}.

\begin{equation}
\dot{M}(Z_i) = \frac{M_{\mathrm{cvz}} X(Z_i)}{t_{\mathrm{set}}(Z_i)}
\label{eq:accretion_rate}
\end{equation}

$M_{\mathrm{cvz}}$ is the mass of the convective envelope, found by multiplying the mass of \WD\ by the convection zone mass ratio \citep{dufour2017}. $X(Z_i)$ is the mass fraction of the element $Z_i$ relative to the dominant element in the atmosphere, and $t_{\mathrm{set}}(Z_i)$ is the settling time for the relevant element ($t_{\mathrm{set}}(\mathrm{Ca})$=701 yr \citep{dufour2017}). Using the measured Ca abundance we calculate the accretion rate to be $\dot{M}_{\text{Ca}}$=$1.52 \times 10^4~\mathrm{g/s}$. We note that this accretion rate is based on a lower limit of the possible mixing in a white dwarf photosphere. If there are extra mixing processes not considered by \cite{koester09}, than the accretion rate could be higher. If we assume the material being accreted onto \WD\ has a composition similar to bulk Earth, the total amount of matter being accreted is $\dot{M}_{\text{Bulk Earth}}$=$9.50\times 10^5~\mathrm{g/s}$.

From this bulk earth accretion rate, we can estimate the total mass of the disk supplying the accreted material. Recent studies have demonstrated that tidally disrupted rocky bodies do not form geometrically flat disks as initially proposed by \citet{jura03}. Instead, the resulting structures may be vertically extended due to the absence of substantial collisional damping \citep{kenyon2017a}. These optically thin, vertically extended disks are typically less massive but still exhibit a large emitting area due to the presence of small dust grains \citep{ballering22}. This scenario aligns with current circumstellar gas observations, which constrain the outer edges of the gas to near the tidal disruption radius as optically thin dust sublimates at greater distances \citep{manser,steele21}. If tidally disrupted bodies relax into quasi-circular swarms that collisionally evolve as presented in \citet{kenyon2017a}, the equilibrium mass ($M_{d,eq}$) for a collisional cascade at the tidal disruption radius ($a$) of a WD would be

\begin{eqnarray}
\label{eqn:diskmass}
M_{d,eq} & \approx & 7\times10^{18}\ \mathrm{g} \left(\frac{\dot{M}}{10^{10} \mathrm{g \cdot s}^{-1}} \right)^\frac{1}{2} \left(\frac{0.6 M_\odot}{M_{\mathrm{WD}}}\right)^\frac{9}{20} \nonumber \\ 
& & \times \left(\frac{r_\mathrm{o}}{1 \mathrm{km}}\right)^{1.04}\left(\frac{\rho}{3.3 \mathrm{g \cdot cm^{-3}}}\right)^\frac{9}{10} \\
& & \times \left(\frac{0.01}{e}\right)^\frac{4}{5}\left(\frac{\Delta a}{0.2a}\right)^\frac{1}{2}\left(\frac{a}{R_{\odot}}\right)^\frac{43}{20}, r_{\mathrm{o}}\ge1~\mathrm{km} \nonumber
\end{eqnarray}

\noindent where $r_{\mathrm o}$ is the characteristic size of the input bodies, $\rho$ is the average density, and $e$ is the eccentricity of the disk. The output gas accretion onto the WD is then equivalent to the influx of mass into the disk in a steady state \citep{kenyon2017b}. Assuming an optically thin, vertically extended, and radially thin debris disk at the Roche limit of the white
dwarf \citep{ballering22}, the equilibrium mass is calculated as $M_{d,eq}$=$2.28\times10^{17}$~g assuming r$_{\mathrm o}$=1~km, $e$=0.1, and $\rho$=3.3~g$\cdot$cm$^{-3}$. In the following section, we compare this disk mass against a photometrically derived disk mass.

\subsection{Dust Disk} \label{subsec:dustdisk}  
\begin{figure*}
\centering
\includegraphics[width=0.75\textwidth]{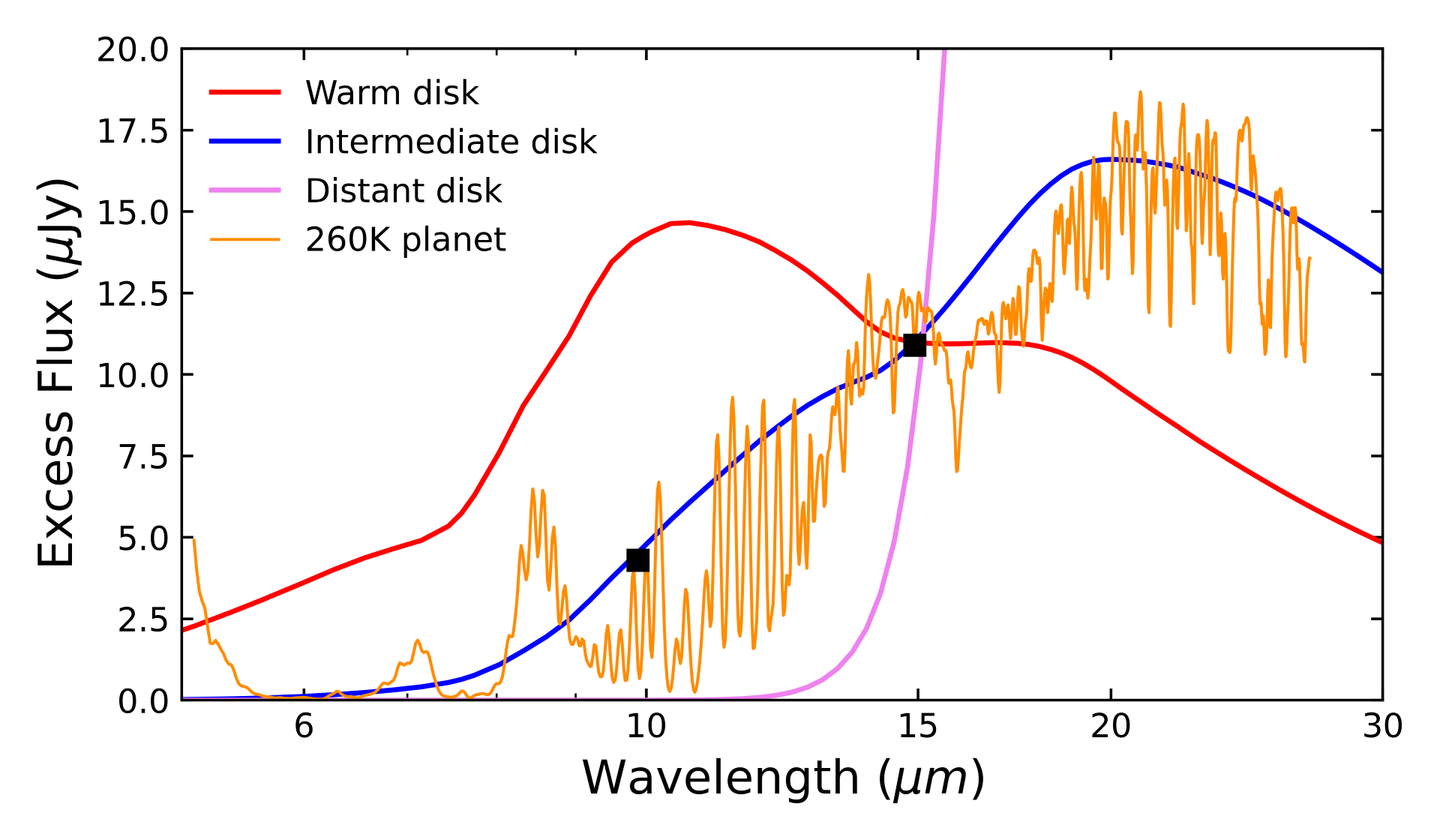}
\caption{Dust disk models fit to 10 and 15 $\mu$m excesses detections for WD 0644+025. The \citet{ballering22} disk model requires $1.1 \times 10^{18}$ g of material in 1 - 10$^{4}$ $\mu m$ particles and is located between 0.86 and 1.29~\Rsun. The intermediate disk model requires $1.3 \times 10^{19}$ g of material in 1 - 10$^{4}$ $\mu m$ particles and is located between 4.3 and 4.7~\Rsun. The distant disk model is located between 645 and 710~\Rsun\ and has a mass of $1.6 \times 10^{28}$ g of material. The planet model describes a 6.8 $M_{Jup}$, 260 K planet at the same age as the white dwarf. }
\label{fig:dustDiskSED}
\end{figure*}

WD 0644+025 exhibits a significant 7.3$\sigma$ excess at $15 \mu$m and fainter 3.6$\sigma$ excess at $10 \mu$m relative to the best fit white dwarf photosphere model, indicating the presence of either a planet or debris disk contributing to the total system emission at longer wavelengths. In this section, we explore the possibility that a debris disk, similar to the debris disks confirmed around other white dwarfs \citep{jura03, kilic06, Jura07} is the source of the excess. To do so, we reproduce the warm debris disk model fit to white dwarf G29-38 presented in \citet{ballering22}, which describes an optically thin, vertically extended, and radially thin debris disk at the Roche limit of the white dwarf, similar in structure to debris disks around main sequence stars. We additionally explore the possibility of an intermediate disk, similar to the distant disk scenario described in \citet{jura08}, and a very distant disk located 10$\times$ further than the \citet{ballering22} disk. Figure \ref{fig:dustDiskSED} shows the 10 and 15~\micron\ fit for the three dust disk models, as well as a 260~K planet model. For a more in-depth description of debris disk models and the fitting procedure, see \cite{debes2025}.

To produce and fit debris disk models to the 10 and 15 $\mu$m photometry points, we model disk thermal emission using radiative transfer code MCFOST \citep{mcfost2, mcfost1}. Given both the uncertainty in white dwarf debris disk morphology and fitting a model to data having only two data points, we opt to fix every disk property except disk mass for all models as to not introduce further uncertainty into our analysis. In each case, we fit the hottest disk possible, but note there are significant degeneracies with disk mass and disk inner radius, so many possible debris disk models may describe the observed excesses.

We find that it is difficult to describe both the 10 and 15~$\mu$m emission with the warm \citet{ballering22} model due to the faint 10 $\mu$m emission relative to 15~$\mu$m emission. In the \citet{ballering22} model, disks exhibit a prominent 10 $\mu$m silicate emission feature, which can only be suppressed by introducing a very large minimum particle size, which would then be inconsistent with dust disk collisional cascade formation theory \citep{kenyon2017b}. The \citet{ballering22} disk model requires $1.1 \times 10^{18}$~g of material in 1 - 10$^{4}$~$\mu m$ particles located at 0.86-1.29~\Rsun\ (0.004-0.006~au), or $3.5 \times 10^{20}$~g of material in 1 - 10$^{9}$~$\mu m$ particles. If we compare this mass to the equilibrium mass of the dust disk predicted by the Ca abundance in \S \ref{subsec:metalAbundance}, we find it is smaller than the equivalent dust disk predicted by photometry by 3 orders of magnitude. While this mass comparison is not certain enough to rule out this model alone, this comparison combined with the overestimated 10~\micron\ feature leads us to conclude that a traditional white dwarf debris disk is unlikely to be the source of the excess emission. 

% Given the temperature of \WD, grain sizes smaller than 1~\micron\ may survive, however a smaller grain size would further amplify the already prominent 10~\micron\ silicate feature, which this model already over predicts.

We find that it is possible to describe both the 10 and 15 $\mu$m emission with a disk similar to the \citet{ballering22} disk, but located at 4.3-4.7~\Rsun\ (0.02-0.022~au). The intermediate disk requires the same order of magnitude of mass as the \citet{ballering22} disk around G29-38, but if dust were to exist there, there would be no theory pointing towards how it got there as tidal disruption is expected to occur closer to the star. The fractional luminosity of this disk would be $\sim$0.20\%, which is an order of magnitude lower than that of many well-studied dusty white dwarfs \citep{becklin2005, reach2005,reach2025}, but still within the range of recently identified faint debris disks, including two of the lowest-luminosity detections reported by \citet{farihi2025}. Empirical constraints on where these dust disks are found are needed if we want to constrain where disruption occurs and through what processes. Although this disk would be atypical relative to the currently known population, our two photometric points are well described by such a model. Follow-up observations are needed to verify the origin of the excess, and a confirmed detection would add a new layer of complexity to current models of white dwarf dust systems. 

%This disk would reside within the radius occupied by the red giant progenitor, and the small particles that constitute the disk would have been removed \citep{jura08}.

Finally, we find the distant disk model is unable to describe both the 10 and 15~$\mu$m emission. Due to the excess emission being at shorter wavelengths as compared to the targets in \cite{debes2025}, far more disk mass is required to describe the excesses. We find that, for a disk spanning 645-710~\Rsun\ (3-3.3~au) and particle size ranging from 1~$\mu$m to $10^{4}~\mu$m, $10^{28}$ g of dust is required. This disk mass would far exceed the mass of the asteroid belt in our solar system, at $2.4\times10^{24}$~g \citep{Pitjeva2018}. Similarly, the distant disk model described in \citet{jura08} presents a conservative upper mass limit of $10^{25}$ g, which is on the order of 1000x less massive than the required disk mass. If we then extrapolate the disk mass out to 300 km sized bodies - the maximum body size presented in the collisional cascade model described in \citet{kenyon2017b}, greater than $10^{31}$g of material is required, exceeding the mass of Jupiter. Given the inability to fit the 10 and 15~\micron\ excesses simultaneously, combined with the unrealistic mass prediction, we conclude that it is likely not the source of the excess. 

\section{Conclusion}

In this work, we present JWST/MIRI observations of the white dwarf \WD, revealing a significant infrared excess at 15~$\mu$m and a marginal excess at 10~$\mu$m. Our modeling indicates that these excesses cannot be easily explained by traditional white dwarf debris disk models, though more unusual disks fit the observed excesses well. The excess is broadly consistent with thermal emission from a substellar companion, and we estimate a best-fit unresolved mass of approximately 6.3~\Mjup, although degeneracies with potential dust emission remain.

We complement this infrared analysis with deep PSF-subtracted imaging and contrast curve analysis, ruling out resolved companions more massive than 2~\Mjup\ at projected separations 12-200~au. Additionally, new Keck HIRES spectroscopy reveals that \WD\ is metal polluted, with an accretion rate consistent with ongoing infall of planetary material— suggesting the presence of at least one surviving planet capable of perturbing minor bodies inward.

Importantly, \WD\ is an unusually massive white dwarf, with a current mass of 0.95~\Msun\ and an inferred progenitor mass of 4.43~\Msun. Stars of this mass undergo rapid and extreme post-main-sequence evolution, with large-scale mass loss and envelope expansion that are expected to destabilize inner planetary systems and reduce the likelihood of planetary survival  \citep{mustill2014}. Theoretical models predict that dynamical instabilities following mass loss from intermediate to high-mass stars often lead to planet–planet scattering and system-wide reconfiguration, with outcomes dominated by ejection rather than inward migration \citep{veras2016}. Nevertheless, a subset of systems may retain giant planets in wide orbits following post-main-sequence mass loss, while others may undergo delayed dynamical instabilities that increase orbital eccentricities and bring surviving planets into much closer orbits around the white dwarf \citep{debes02, veras2013}. Furthermore, recent work by \cite{limbach2025} confirmed the first intact exoplanet within a region where planets would have been engulfed during the red giant phase- providing direct evidence that planetary migration into close orbits is possible. The confirmation of another close-in companion would place stronger constraints on the dynamical pathways by which such companions can survive or migrate inward after the host star’s evolution.

Together, these findings point toward either a giant planet or an unusual dust structure contributing to the observed infrared excess of \WD. The Gaia DR3 catalog reports a RUWE of 1.065 for \WD, indicating that the single-star astrometric solution is well behaved and showing no evidence for a large perturbation at the current precision. While a value near 1.06 is not anomalous, it does not rule out the presence of a companion at separations of a few au. Using the expected DR4 astrometric performance and the scaling relations from \citet{Sanderson2022}, we estimate that the DR4 astrometric signal-to-noise for a $\sim$6.8~\Mjup\ companion increases approximately as ${\rm S/N} \simeq 2.5\,(a/{\rm au})$, suggesting that DR4 should be sensitive to such a planet over a wide range of separations. Gaia is most sensitive to orbits with periods shorter than the mission duration. For DR4, the 5.5-year mission length corresponds to a semi-major axis of $a \simeq 3.08$~au for this system. However, initial results from Gaia DR3 comparing spectroscopically and astrometrically determined orbital periods suggest that periods up to $\sim$20-40\% longer than the mission length may still be recoverable \citep{Pourbaix2022}, implying that partially sampled orbits at somewhat larger separations could still produce measurable astrometric trends. Mid-infrared spectroscopy with JWST/MIRI would also help discriminate between these scenarios by revealing molecular absorption features characteristic of a cool giant planet or, alternatively, the continuum expected from thermal dust. Continued monitoring—both spectroscopic and astrometric—will therefore be essential to distinguish between a planetary companion and a dust disk, and to clarify the architecture and evolutionary pathway of this white dwarf system.

\clearpage
\onecolumngrid

\begin{acknowledgments}
We acknowledge support by the NSF under grant AST-2205736 and AST-2508429, and NASA under grant Nos. 80NSSC22K0479, 80NSSC24K0380, and 80NSSC24K0436.

This work is based on observations with the NASA/ESA/CSA James Webb Space Telescope obtained from the Mikulski Archive for Space Telescopes at the Space Telescope Science Institute, which is operated by the Association of Universities for Research in Astronomy, Incorporated, under NASA contract NAS 5-03127. These observations are associated with program \#3964. 

This work was supported by a NASA Keck PI Data Award, administered by the NASA Exoplanet Science Institute. Data presented herein were obtained at the W.M. Keck Observatory from telescope time allocated to the National Aeronautics and Space Administration through the agency scientific partnership with the California Institute of Technology and the University of California. The Observatory was made possible by the generous financial support of the W. M. Keck Foundation.

The authors wish to recognize and acknowledge the very significant cultural role and reverence that the summit of Maunakea has always had within the indigenous Hawaiian community. We are most fortunate to have the opportunity to conduct observations from this mountain.

Funding for the Sloan Digital Sky Survey V has been provided by the Alfred P. Sloan Foundation, the Heising-Simons Foundation, the National Science Foundation, and the Participating Institutions. SDSS acknowledges support and resources from the Center for High-Performance Computing at the University of Utah. SDSS telescopes are located at Apache Point Observatory, funded by the Astrophysical Research Consortium and operated by New Mexico State University, and at Las Campanas Observatory, operated by the Carnegie Institution for Science. The SDSS web site is \url{www.sdss.org}.

SDSS is managed by the Astrophysical Research Consortium for the Participating Institutions of the SDSS Collaboration, including the Carnegie Institution for Science, Chilean National Time Allocation Committee (CNTAC) ratified researchers, Caltech, the Gotham Participation Group, Harvard University, Heidelberg University, The Flatiron Institute, The Johns Hopkins University, L'Ecole polytechnique f\'{e}d\'{e}rale de Lausanne (EPFL), Leibniz-Institut f\"{u}r Astrophysik Potsdam (AIP), Max-Planck-Institut f\"{u}r Astronomie (MPIA Heidelberg), Max-Planck-Institut f\"{u}r Extraterrestrische Physik (MPE), Nanjing University, National Astronomical Observatories of China (NAOC), New Mexico State University, The Ohio State University, Pennsylvania State University, Smithsonian Astrophysical Observatory, Space Telescope Science Institute (STScI), the Stellar Astrophysics Participation Group, Universidad Nacional Aut\'{o}noma de M\'{e}xico, University of Arizona, University of Colorado Boulder, University of Illinois at Urbana-Champaign, University of Toronto, University of Utah, University of Virginia, Yale University, and Yunnan University. 

L.A. warmly thanks the Canadian Space Agency for its financial support through grant 23JWGO2B07.

\end{acknowledgments}






\bibliography{main}{}

@ARTICLE{steele21,
       author = {{Steele}, Amy and {Debes}, John and {Xu}, Siyi and {Yeh}, Sherry and {Dufour}, Patrick},
        title = "{A Characterization of the Circumstellar Gas of WD 1124-293 Using Cloudy}",
      journal = {\apj},
         year = 2021,
        month = apr,
       volume = {911},
       number = {1},
          eid = {25},
        pages = {25},
          doi = {10.3847/1538-4357/abc262},
archivePrefix = {arXiv},
       eprint = {2010.12667},
 primaryClass = {astro-ph.EP},
       adsurl = {https://ui.adsabs.harvard.edu/abs/2021ApJ...911...25S}
}

@ARTICLE{manser,
       author = {{Manser}, Christopher J. and {G{\"a}nsicke}, Boris T. and {Marsh}, Thomas R. and {Veras}, Dimitri and {Koester}, Detlev and {Breedt}, Elm{\'e} and {Pala}, Anna F. and {Parsons}, Steven G. and {Southworth}, John},
        title = "{Doppler imaging of the planetary debris disc at the white dwarf SDSS J122859.93+104032.9}",
      journal = {\mnras},
           year = 2016,
        month = feb,
       volume = {455},
       number = {4},
        pages = {4467-4478},
          doi = {10.1093/mnras/stv2603},
archivePrefix = {arXiv},
       eprint = {1511.02230},
 primaryClass = {astro-ph.SR},
       adsurl = {https://ui.adsabs.harvard.edu/abs/2016MNRAS.455.4467M}
}

@ARTICLE{ballering22,
       author = {{Ballering}, Nicholas P. and {Levens}, Colette I. and {Su}, Kate Y.~L. and {Cleeves}, L. Ilsedore},
        title = "{The Geometry of the G29-38 White Dwarf Dust Disk from Radiative Transfer Modeling}",
      journal = {\apj},
         year = 2022,
        month = nov,
       volume = {939},
       number = {2},
          eid = {108},
        pages = {108},
          doi = {10.3847/1538-4357/ac9a4a},
archivePrefix = {arXiv},
       eprint = {2211.00118},
 primaryClass = {astro-ph.EP},
       adsurl = {https://ui.adsabs.harvard.edu/abs/2022ApJ...939..108B}
}

@misc{SonoraBob,
      author = {{Marley}, Mark and {Saumon}, Didier and {Morley}, Caroline and {Fortney}, Jonathan and {Visscher}, Channon and {Freedman}, Richard and {Lupu}, Roxana},
      title = {{Sonora Bobcat: cloud-free, substellar atmosphere models, spectra, photometry, evolution, and chemistry [Data set]}},
      month = jul,
      year = 2021,
      doi = {10.5281/zenodo.5063476},
      version = {Sonora Bobcat},
      publisher = {Zenodo},
      url = {https://doi.org/10.5281/zenodo.5063476}
      }

@ARTICLE{bedard20,
       author = {{B{\'e}dard}, A. and {Bergeron}, P. and {Brassard}, P. and {Fontaine}, G.},
        title = "{On the Spectral Evolution of Hot White Dwarf Stars. I. A Detailed Model Atmosphere Analysis of Hot White Dwarfs from SDSS DR12}",
      journal = {\apj},
         year = 2020,
        month = oct,
       volume = {901},
       number = {2},
          eid = {93},
        pages = {93},
          doi = {10.3847/1538-4357/abafbe},
archivePrefix = {arXiv},
       eprint = {2008.07469},
 primaryClass = {astro-ph.SR},
       adsurl = {https://ui.adsabs.harvard.edu/abs/2020ApJ...901...93B}
}

@ARTICLE{holberg06,
       author = {{Holberg}, J.~B. and {Bergeron}, Pierre},
        title = "{Calibration of Synthetic Photometry Using DA White Dwarfs}",
      journal = {\aj},
         year = 2006,
        month = sep,
       volume = {132},
       number = {3},
        pages = {1221-1233},
          doi = {10.1086/505938},
       adsurl = {https://ui.adsabs.harvard.edu/abs/2006AJ....132.1221H}
}

@ARTICLE{linder19,
       author = {{Linder}, Esther F. and {Mordasini}, Christoph and {Molli{\`e}re}, Paul and {Marleau}, Gabriel-Dominique and {Malik}, Matej and {Quanz}, Sascha P. and {Meyer}, Michael R.},
        title = "{Evolutionary models of cold and low-mass planets: cooling curves, magnitudes, and detectability}",
      journal = {\aap},
         year = 2019,
        month = mar,
       volume = {623},
          eid = {A85},
        pages = {A85},
          doi = {10.1051/0004-6361/201833873},
archivePrefix = {arXiv},
       eprint = {1812.02027},
 primaryClass = {astro-ph.EP},
       adsurl = {https://ui.adsabs.harvard.edu/abs/2019A&A...623A..85L}
}

@ARTICLE{kenyon2017a,
       author = {{Kenyon}, Scott J. and {Bromley}, Benjamin C.},
        title = "{Numerical Simulations of Collisional Cascades at the Roche Limits of White Dwarf Stars}",
      journal = {\apj},
         year = 2017,
        month = aug,
       volume = {844},
       number = {2},
          eid = {116},
        pages = {116},
          doi = {10.3847/1538-4357/aa7b85},
archivePrefix = {arXiv},
       eprint = {1706.08579},
 primaryClass = {astro-ph.SR},
       adsurl = {https://ui.adsabs.harvard.edu/abs/2017ApJ...844..116K}
}

@ARTICLE{debes05,
       author = {{Debes}, John H. and {Sigurdsson}, Steinn and {Woodgate}, Bruce E.},
        title = "{Cool Customers in the Stellar Graveyard. II. Limits to Substellar Objects around Nearby DAZ White Dwarfs}",
      journal = {\aj},
         year = 2005,
        month = sep,
       volume = {130},
       number = {3},
        pages = {1221-1230},
          doi = {10.1086/432660},
archivePrefix = {arXiv},
       eprint = {astro-ph/0412669},
 primaryClass = {astro-ph},
       adsurl = {https://ui.adsabs.harvard.edu/abs/2005AJ....130.1221D}
}

@ARTICLE{mullally07,
       author = {{Mullally}, F. and {Kilic}, Mukremin and {Reach}, William T. and {Kuchner}, Marc J. and {von Hippel}, Ted and {Burrows}, Adam and {Winget}, D.~E.},
        title = "{A Spitzer White Dwarf Infrared Survey}",
      journal = {\apjs},
         year = 2007,
        month = jul,
       volume = {171},
       number = {1},
        pages = {206-218},
          doi = {10.1086/511858},
archivePrefix = {arXiv},
       eprint = {astro-ph/0611588},
 primaryClass = {astro-ph},
       adsurl = {https://ui.adsabs.harvard.edu/abs/2007ApJS..171..206M}
}

@ARTICLE{brandner21,
       author = {{Brandner}, Wolfgang and {Zinnecker}, Hans and {Kopytova}, Taisiya},
        title = "{Search for giant planets around seven white dwarfs in the Hyades cluster with the Hubble Space Telescope}",
      journal = {\mnras},
         year = 2021,
        month = jan,
       volume = {500},
       number = {3},
        pages = {3920-3925},
          doi = {10.1093/mnras/staa3422},
archivePrefix = {arXiv},
       eprint = {2011.03562},
 primaryClass = {astro-ph.EP},
       adsurl = {https://ui.adsabs.harvard.edu/abs/2021MNRAS.500.3920B}
}

@ARTICLE{luhman12,
       author = {{Luhman}, K.~L. and {Burgasser}, A.~J. and {Labb{\'e}}, I. and {Saumon}, D. and {Marley}, M.~S. and {Bochanski}, J.~J. and {Monson}, A.~J. and {Persson}, S.~E.},
        title = "{Confirmation of One of the Coldest Known Brown Dwarfs}",
      journal = {\apj},
         year = 2012,
        month = jan,
       volume = {744},
       number = {2},
          eid = {135},
        pages = {135},
          doi = {10.1088/0004-637X/744/2/135},
archivePrefix = {arXiv},
       eprint = {1110.4353},
 primaryClass = {astro-ph.GA},
       adsurl = {https://ui.adsabs.harvard.edu/abs/2012ApJ...744..135L}
}

@ARTICLE{koester09,
       author = {{Koester}, D.},
        title = "{Accretion and diffusion in white dwarfs. New diffusion timescales and applications to GD 362 and G 29-38}",
      journal = {\aap},
         year = 2009,
        month = may,
       volume = {498},
       number = {2},
        pages = {517-525},
          doi = {10.1051/0004-6361/200811468},
archivePrefix = {arXiv},
       eprint = {0903.1499},
 primaryClass = {astro-ph.SR},
       adsurl = {https://ui.adsabs.harvard.edu/abs/2009A&A...498..517K}
}

@ARTICLE{zuckerman03,
       author = {{Zuckerman}, B. and {Koester}, D. and {Reid}, I.~N. and {H{\"u}nsch}, M.},
        title = "{Metal Lines in DA White Dwarfs}",
      journal = {\apj},
         year = 2003,
        month = oct,
       volume = {596},
       number = {1},
        pages = {477-495},
          doi = {10.1086/377492},
       adsurl = {https://ui.adsabs.harvard.edu/abs/2003ApJ...596..477Z}
}

@ARTICLE{alcock86,
       author = {{Alcock}, C. and {Fristrom}, C.~C. and {Siegelman}, R.},
        title = "{On the Number of Comets around Other Single Stars}",
      journal = {\apj},
         year = 1986,
        month = mar,
       volume = {302},
        pages = {462},
          doi = {10.1086/164005},
       adsurl = {https://ui.adsabs.harvard.edu/abs/1986ApJ...302..462A}
}

@ARTICLE{jura03,
       author = {{Jura}, M.},
        title = "{A Tidally Disrupted Asteroid around the White Dwarf G29-38}",
      journal = {\apjl},
         year = 2003,
        month = feb,
       volume = {584},
       number = {2},
        pages = {L91-L94},
          doi = {10.1086/374036},
archivePrefix = {arXiv},
       eprint = {astro-ph/0301411},
 primaryClass = {astro-ph},
       adsurl = {https://ui.adsabs.harvard.edu/abs/2003ApJ...584L..91J}
}

@ARTICLE{kiman22,
       author = {{Kiman}, Rocio and {Xu}, Siyi and {Faherty}, Jacqueline K. and {Gagn{\'e}}, Jonathan and {Angus}, Ruth and {Brandt}, Timothy D. and {Casewell}, Sarah L. and {Cruz}, Kelle L.},
        title = "{wdwarfdate: A Python Package to Derive Bayesian Ages of White Dwarfs}",
      journal = {\aj},
         year = 2022,
        month = aug,
       volume = {164},
       number = {2},
          eid = {62},
        pages = {62},
          doi = {10.3847/1538-3881/ac7788},
archivePrefix = {arXiv},
       eprint = {2206.05388},
 primaryClass = {astro-ph.SR},
       adsurl = {https://ui.adsabs.harvard.edu/abs/2022AJ....164...62K}
}

@ARTICLE{cummings18,
       author = {{Cummings}, Jeffrey D. and {Kalirai}, Jason S. and {Tremblay}, P. -E. and {Ramirez-Ruiz}, Enrico and {Choi}, Jieun},
        title = "{The White Dwarf Initial-Final Mass Relation for Progenitor Stars from 0.85 to 7.5 M $_{{\ensuremath{\odot}}}$}",
      journal = {\apj},
         year = 2018,
        month = oct,
       volume = {866},
       number = {1},
          eid = {21},
        pages = {21},
          doi = {10.3847/1538-4357/aadfd6},
archivePrefix = {arXiv},
       eprint = {1809.01673},
 primaryClass = {astro-ph.SR},
       adsurl = {https://ui.adsabs.harvard.edu/abs/2018ApJ...866...21C}
}

@ARTICLE{choi16,
       author = {{Choi}, Jieun and {Dotter}, Aaron and {Conroy}, Charlie and {Cantiello}, Matteo and {Paxton}, Bill and {Johnson}, Benjamin D.},
        title = "{Mesa Isochrones and Stellar Tracks (MIST). I. Solar-scaled Models}",
      journal = {\apj},
         year = 2016,
        month = jun,
       volume = {823},
       number = {2},
          eid = {102},
        pages = {102},
          doi = {10.3847/0004-637X/823/2/102},
archivePrefix = {arXiv},
       eprint = {1604.08592},
 primaryClass = {astro-ph.SR},
       adsurl = {https://ui.adsabs.harvard.edu/abs/2016ApJ...823..102C}
}

@ARTICLE{dotter16,
       author = {{Dotter}, Aaron},
        title = "{MESA Isochrones and Stellar Tracks (MIST) 0: Methods for the Construction of Stellar Isochrones}",
      journal = {\apjs},
         year = 2016,
        month = jan,
       volume = {222},
       number = {1},
          eid = {8},
        pages = {8},
          doi = {10.3847/0067-0049/222/1/8},
archivePrefix = {arXiv},
       eprint = {1601.05144},
 primaryClass = {astro-ph.SR},
       adsurl = {https://ui.adsabs.harvard.edu/abs/2016ApJS..222....8D}
}

@ARTICLE{debes02,
       author = {{Debes}, John H. and {Sigurdsson}, Steinn},
        title = "{Are There Unstable Planetary Systems around White Dwarfs?}",
      journal = {\apj},
         year = 2002,
        month = jun,
       volume = {572},
       number = {1},
        pages = {556-565},
          doi = {10.1086/340291},
archivePrefix = {arXiv},
       eprint = {astro-ph/0202273},
 primaryClass = {astro-ph},
       adsurl = {https://ui.adsabs.harvard.edu/abs/2002ApJ...572..556D}
}

@ARTICLE{jura08,
       author = {{Jura}, M.},
        title = "{Pollution of Single White Dwarfs by Accretion of Many Small Asteroids}",
      journal = {\aj},
         year = 2008,
        month = may,
       volume = {135},
       number = {5},
        pages = {1785-1792},
          doi = {10.1088/0004-6256/135/5/1785},
archivePrefix = {arXiv},
       eprint = {0802.4075},
 primaryClass = {astro-ph},
       adsurl = {https://ui.adsabs.harvard.edu/abs/2008AJ....135.1785J}
}

@ARTICLE{dupuis93,
       author = {{Dupuis}, J. and {Fontaine}, G. and {Wesemael}, F.},
        title = "{A Study of Metal Abundance Patterns in Cool White Dwarfs. III. Comparison of the Predictions of the Two-Phase Accretion Model with the Observations}",
      journal = {\apjs},
         year = 1993,
        month = jul,
       volume = {87},
        pages = {345},
          doi = {10.1086/191808},
       adsurl = {https://ui.adsabs.harvard.edu/abs/1993ApJS...87..345D}
}

@ARTICLE{dupuis92,
       author = {{Dupuis}, J. and {Fontaine}, G. and {Pelletier}, C. and {Wesemael}, F.},
        title = "{A Study of Metal Abundance Patterns in Cool White Dwarfs. I. Time-dependent Calculations of Gravitational Settling}",
      journal = {\apjs},
         year = 1992,
        month = oct,
       volume = {82},
        pages = {505},
          doi = {10.1086/191728},
       adsurl = {https://ui.adsabs.harvard.edu/abs/1992ApJS...82..505D}
}

@ARTICLE{hansen03,
       author = {{Hansen}, Brad M.~S. and {Liebert}, James},
        title = "{Cool White Dwarfs}",
      journal = {\araa},
         year = 2003,
        month = jan,
       volume = {41},
        pages = {465-515},
          doi = {10.1146/annurev.astro.41.081401.155117},
       adsurl = {https://ui.adsabs.harvard.edu/abs/2003ARA&A..41..465H}
}

@INPROCEEDINGS{jura14,
       author = {{Jura}, M.},
        title = "{The Elemental Compositions of Extrasolar Planetesimals}",
    booktitle = {Formation, Detection, and Characterization of Extrasolar Habitable Planets},
         year = 2014,
       editor = {{Haghighipour}, Nader},
       volume = {293},
        month = apr,
        pages = {219-228},
          doi = {10.1017/S1743921313012878},
archivePrefix = {arXiv},
       eprint = {1301.5562},
 primaryClass = {astro-ph.EP},
       adsurl = {https://ui.adsabs.harvard.edu/abs/2014IAUS..293..219J}
}

@ARTICLE{wolthoff22,
       author = {{Wolthoff}, Vera and {Reffert}, Sabine and {Quirrenbach}, Andreas and {Jones}, Mat{\'\i}as I. and {Wittenmyer}, Robert A. and {Jenkins}, James S.},
        title = "{Precise radial velocities of giant stars. XVI. Planet occurrence rates from the combined analysis of the Lick, EXPRESS, and PPPS giant star surveys}",
      journal = {\aap},
         year = 2022,
        month = may,
       volume = {661},
          eid = {A63},
        pages = {A63},
          doi = {10.1051/0004-6361/202142501},
archivePrefix = {arXiv},
       eprint = {2202.12800},
 primaryClass = {astro-ph.EP},
       adsurl = {https://ui.adsabs.harvard.edu/abs/2022A&A...661A..63W}
}

@ARTICLE{mustill12,
       author = {{Mustill}, Alexander J. and {Villaver}, Eva},
        title = "{Foretellings of Ragnar{\"o}k: World-engulfing Asymptotic Giants and the Inheritance of White Dwarfs}",
      journal = {\apj},
         year = 2012,
        month = dec,
       volume = {761},
       number = {2},
          eid = {121},
        pages = {121},
          doi = {10.1088/0004-637X/761/2/121},
archivePrefix = {arXiv},
       eprint = {1210.0328},
 primaryClass = {astro-ph.EP},
       adsurl = {https://ui.adsabs.harvard.edu/abs/2012ApJ...761..121M}
}

@ARTICLE{koester14,
       author = {{Koester}, D. and {G{\"a}nsicke}, B.~T. and {Farihi}, J.},
        title = "{The frequency of planetary debris around young white dwarfs}",
      journal = {\aap},
         year = 2014,
        month = jun,
       volume = {566},
          eid = {A34},
        pages = {A34},
          doi = {10.1051/0004-6361/201423691},
archivePrefix = {arXiv},
       eprint = {1404.2617},
 primaryClass = {astro-ph.SR},
       adsurl = {https://ui.adsabs.harvard.edu/abs/2014A&A...566A..34K}
}

@ARTICLE{kilic09,
       author = {{Kilic}, Mukremin and {Gould}, Andrew and {Koester}, Detlev},
        title = "{Limits on Unresolved Planetary Companions to White Dwarf Remnants of 14 Intermediate-Mass Stars}",
      journal = {\apj},
         year = 2009,
        month = nov,
       volume = {705},
       number = {2},
        pages = {1219-1225},
          doi = {10.1088/0004-637X/705/2/1219},
archivePrefix = {arXiv},
       eprint = {0909.2038},
 primaryClass = {astro-ph.SR},
       adsurl = {https://ui.adsabs.harvard.edu/abs/2009ApJ...705.1219K}
}

@ARTICLE{blackman21,
       author = {{Blackman}, J.~W. and {Beaulieu}, J.~P. and {Bennett}, D.~P. and {Danielski}, C. and {Alard}, C. and {Cole}, A.~A. and {Vandorou}, A. and {Ranc}, C. and {Terry}, S.~K. and {Bhattacharya}, A. and {Bond}, I. and {Bachelet}, E. and {Veras}, D. and {Koshimoto}, N. and {Batista}, V. and {Marquette}, J.~B.},
        title = "{A Jovian analogue orbiting a white dwarf star}",
      journal = {\nat},
         year = 2021,
        month = oct,
       volume = {598},
       number = {7880},
        pages = {272-275},
          doi = {10.1038/s41586-021-03869-6},
archivePrefix = {arXiv},
       eprint = {2110.07934},
 primaryClass = {astro-ph.EP},
       adsurl = {https://ui.adsabs.harvard.edu/abs/2021Natur.598..272B}
}

@ARTICLE{vanderburg2020,
       author = {{Vanderburg}, Andrew and {Rappaport}, Saul A. and {Xu}, Siyi and
         {Crossfield}, Ian J.~M. and {Becker}, Juliette C. and {Gary}, Bruce and
         {Murgas}, Felipe and {Blouin}, Simon and {Kaye}, Thomas G. and
         {Palle}, Enric and {Melis}, Carl and {Morris}, Brett M. and
         {Kreidberg}, Laura and {Gorjian}, Varoujan and {Morley}, Caroline V. and
         {Mann}, Andrew W. and {Parviainen}, Hannu and {Pearce}, Logan A. and
         {Newton}, Elisabeth R. and {Carrillo}, Andreia and {Zuckerman}, Ben and
         {Nelson}, Lorne and {Zeimann}, Greg and {Brown}, Warren R. and
         {Tronsgaard}, Ren{\'e} and {Klein}, Beth and {Ricker}, George R. and {Vand
        erspek}, Roland K. and {Latham}, David W. and {Seager}, Sara and
         {Winn}, Joshua N. and {Jenkins}, Jon M. and {Adams}, Fred C. and
         {Benneke}, Bj{\"o}rn and {Berardo}, David and {Buchhave}, Lars A. and
         {Caldwell}, Douglas A. and {Christiansen}, Jessie L. and
         {Collins}, Karen A. and {Col{\'o}n}, Knicole D. and {Daylan}, Tansu and
         {Doty}, John and {Doyle}, Alexandra E. and {Dragomir}, Diana and
         {Dressing}, Courtney and {Dufour}, Patrick and {Fukui}, Akihiko and
         {Glidden}, Ana and {Guerrero}, Natalia M. and {Guo}, Xueying and
         {Heng}, Kevin and {Henriksen}, Andreea I. and {Huang}, Chelsea X. and
         {Kaltenegger}, Lisa and {Kane}, Stephen R. and {Lewis}, John A. and
         {Lissauer}, Jack J. and {Morales}, Farisa and {Narita}, Norio and
         {Pepper}, Joshua and {Rose}, Mark E. and {Smith}, Jeffrey C. and
         {Stassun}, Keivan G. and {Yu}, Liang},
        title = "{A giant planet candidate transiting a white dwarf}",
      journal = {\nat},
         year = 2020,
        month = sep,
       volume = {585},
       number = {7825},
        pages = {363-367},
          doi = {10.1038/s41586-020-2713-y},
archivePrefix = {arXiv},
       eprint = {2009.07282},
 primaryClass = {astro-ph.EP},
       adsurl = {https://ui.adsabs.harvard.edu/abs/2020Natur.585..363V}
}

@MISC{jdox16,
        title = "{JWST User Documentation (JDox)}",
    author={{JDox}},
 howpublished = {JWST User Documentation Website},
         year = 2016,
        month = jan,
       adsurl = {https://ui.adsabs.harvard.edu/abs/2016jdox.rept......}
}

@ARTICLE{mullally24,
       author = {{Mullally}, Susan E. and {Debes}, John and {Cracraft}, Misty and {Mullally}, Fergal and {Poulsen}, Sabrina and {Albert}, Loic and {Thibault}, Katherine and {Reach}, William T. and {Hermes}, J.~J. and {Barclay}, Thomas and {Kilic}, Mukremin and {Quintana}, Elisa V.},
        title = "{JWST Directly Images Giant Planet Candidates Around Two Metal-Polluted White Dwarf Stars}",
      journal = {arXiv e-prints},
         year = 2024,
        month = jan,
          eid = {arXiv:2401.13153},
        pages = {arXiv:2401.13153},
          doi = {10.48550/arXiv.2401.13153},
archivePrefix = {arXiv},
       eprint = {2401.13153},
 primaryClass = {astro-ph.EP},
       adsurl = {https://ui.adsabs.harvard.edu/abs/2024arXiv240113153M}
}

@ARTICLE{hagai10,
       author = {{Perets}, Hagai B.},
        title = "{Second generation planets}",
      journal = {arXiv e-prints},
         year = 2010,
        month = jan,
          eid = {arXiv:1001.0581},
        pages = {arXiv:1001.0581},
          doi = {10.48550/arXiv.1001.0581},
archivePrefix = {arXiv},
       eprint = {1001.0581},
 primaryClass = {astro-ph.EP},
       adsurl = {https://ui.adsabs.harvard.edu/abs/2010arXiv1001.0581P}
}

@ARTICLE{mcfost1,
       author = {{Pinte}, C. and {Harries}, T.~J. and {Min}, M. and {Watson}, A.~M. and {Dullemond}, C.~P. and {Woitke}, P. and {M{\'e}nard}, F. and {Dur{\'a}n-Rojas}, M.~C.},
        title = "{Benchmark problems for continuum radiative transfer. High optical depths, anisotropic scattering, and polarisation}",
      journal = {\aap},
         year = 2009,
        month = may,
       volume = {498},
       number = {3},
        pages = {967-980},
          doi = {10.1051/0004-6361/200811555},
archivePrefix = {arXiv},
       eprint = {0903.1231},
 primaryClass = {astro-ph.SR},
       adsurl = {https://ui.adsabs.harvard.edu/abs/2009A&A...498..967P}
}

@ARTICLE{mcfost2,
       author = {{Pinte}, C. and {M{\'e}nard}, F. and {Duch{\^e}ne}, G. and {Bastien}, P.},
        title = "{Monte Carlo radiative transfer in protoplanetary disks}",
      journal = {\aap},
         year = 2006,
        month = dec,
       volume = {459},
       number = {3},
        pages = {797-804},
          doi = {10.1051/0004-6361:20053275},
archivePrefix = {arXiv},
       eprint = {astro-ph/0606550},
 primaryClass = {astro-ph},
       adsurl = {https://ui.adsabs.harvard.edu/abs/2006A&A...459..797P}
}

@ARTICLE{kenyon2017b,
       author = {{Kenyon}, Scott J. and {Bromley}, Benjamin C.},
        title = "{Numerical Simulations of Gaseous Disks Generated from Collisional Cascades at the Roche Limits of White Dwarf Stars}",
      journal = {\apj},
         year = 2017,
        month = nov,
       volume = {850},
       number = {1},
          eid = {50},
        pages = {50},
          doi = {10.3847/1538-4357/aa9570},
archivePrefix = {arXiv},
       eprint = {1711.00026},
 primaryClass = {astro-ph.SR},
       adsurl = {https://ui.adsabs.harvard.edu/abs/2017ApJ...850...50K}
}

@article{Pitjeva2018,
   title={Masses of the Main Asteroid Belt and the Kuiper Belt from the Motions of Planets and Spacecraft},
   volume={44},
   ISSN={1562-6873},
   url={http://dx.doi.org/10.1134/S1063773718090050},
   DOI={10.1134/s1063773718090050},
   number={8–9},
   journal={Astronomy Letters},
   publisher={Pleiades Publishing Ltd},
   author={Pitjeva, E. V. and Pitjev, N. P.},
   year={2018},
   month=aug, pages={554–566} }

@ARTICLE{kilic06,
       author = {{Kilic}, Mukremin and {von Hippel}, Ted and {Leggett}, S.~K. and {Winget}, D.~E.},
        title = "{Debris Disks around White Dwarfs: The DAZ Connection}",
      journal = {\apj},
         year = 2006,
        month = jul,
       volume = {646},
       number = {1},
        pages = {474-479},
          doi = {10.1086/504682},
archivePrefix = {arXiv},
       eprint = {astro-ph/0603774},
 primaryClass = {astro-ph},
       adsurl = {https://ui.adsabs.harvard.edu/abs/2006ApJ...646..474K}
}

@ARTICLE{Jura07,
       author = {{Jura}, M. and {Farihi}, J. and {Zuckerman}, B.},
        title = "{Externally Polluted White Dwarfs with Dust Disks}",
      journal = {\apj},
         year = 2007,
        month = jul,
       volume = {663},
       number = {2},
        pages = {1285-1290},
          doi = {10.1086/518767},
archivePrefix = {arXiv},
       eprint = {0704.1170},
 primaryClass = {astro-ph},
       adsurl = {https://ui.adsabs.harvard.edu/abs/2007ApJ...663.1285J}
}

@ARTICLE{poulsen24,
       author = {{Poulsen}, Sabrina and {Debes}, John and {Cracraft}, Misty and {Mullally}, Susan E. and {Reach}, William T. and {Kilic}, Mukremin and {Mullally}, Fergal and {Albert}, Loic and {Thibault}, Katherine and {Hermes}, J.~J. and {Barclay}, Thomas and {Quintana}, Elisa V.},
        title = "{A MIRI Search for Planets and Dust around WD 2149+021}",
      journal = {\aj},
         year = 2024,
        month = jun,
       volume = {167},
       number = {6},
          eid = {257},
        pages = {257},
          doi = {10.3847/1538-3881/ad374c},
archivePrefix = {arXiv},
       eprint = {2311.14165},
 primaryClass = {astro-ph.EP},
       adsurl = {https://ui.adsabs.harvard.edu/abs/2024AJ....167..257P}
}

@ARTICLE{hintzen1974,
       author = {{Hintzen}, Paul and {Strittmatter}, P.~A.},
        title = "{A Spectroscopic Search for Cool White Dwarfs}",
      journal = {\apjl},
         year = 1974,
        month = nov,
       volume = {193},
        pages = {L111},
          doi = {10.1086/181645},
       adsurl = {https://ui.adsabs.harvard.edu/abs/1974ApJ...193L.111H}
}

@ARTICLE{bagnulo2021,
       author = {{Bagnulo}, S. and {Landstreet}, J.~D.},
        title = "{New insight into the magnetism of degenerate stars from the analysis of a volume-limited sample of white dwarfs}",
      journal = {\mnras},
         year = 2021,
        month = nov,
       volume = {507},
       number = {4},
        pages = {5902-5951},
          doi = {10.1093/mnras/stab2046},
archivePrefix = {arXiv},
       eprint = {2106.11109},
 primaryClass = {astro-ph.SR},
       adsurl = {https://ui.adsabs.harvard.edu/abs/2021MNRAS.507.5902B}
}

@ARTICLE{barber2016,
       author = {{Barber}, Sara D. and {Belardi}, Claudia and {Kilic}, Mukremin and {Gianninas}, A.},
        title = "{Remnant planetary systems around bright white dwarfs}",
      journal = {\mnras},
         year = 2016,
        month = jun,
       volume = {459},
       number = {2},
        pages = {1415-1421},
          doi = {10.1093/mnras/stw683},
       adsurl = {https://ui.adsabs.harvard.edu/abs/2016MNRAS.459.1415B}
}

@article{zuckerman2003,
doi = {10.1086/377492},
url = {https://dx.doi.org/10.1086/377492},
year = {2003},
month = {oct},
publisher = {},
volume = {596},
number = {1},
pages = {477},
author = {Zuckerman, B. and Koester, D. and Reid, I. N. and Hünsch, M.},
title = {Metal Lines in DA White Dwarfs*},
journal = {The Astrophysical Journal}
}

@ARTICLE{dufour2012,
       author = {{Dufour}, P. and {Kilic}, M. and {Fontaine}, G. and {Bergeron}, P. and {Melis}, C. and {Bochanski}, J.},
        title = "{Detailed Compositional Analysis of the Heavily Polluted DBZ White Dwarf SDSS J073842.56+183509.06: A Window on Planet Formation?}",
      journal = {\apj},
         year = 2012,
        month = apr,
       volume = {749},
       number = {1},
          eid = {6},
        pages = {6},
          doi = {10.1088/0004-637X/749/1/6},
archivePrefix = {arXiv},
       eprint = {1201.6252},
 primaryClass = {astro-ph.SR},
       adsurl = {https://ui.adsabs.harvard.edu/abs/2012ApJ...749....6D}
}

@ARTICLE{limbach2024,
       author = {{Limbach}, Mary Anne and {Vanderburg}, Andrew and {Venner}, Alexander and {Blouin}, Simon and {Stevenson}, Kevin B. and {MacDonald}, Ryan J. and {Jenkins}, Sydney and {Bowens-Rubin}, Rachel and {Soares-Furtado}, Melinda and {Morley}, Caroline and {Janson}, Markus and {Debes}, John and {Xu}, Siyi and {Kleisioti}, Evangelia and {Kenworthy}, Matthew and {Butler}, Paul and {Crane}, Jeffrey D. and {Osip}, Dave and {Shectman}, Stephen and {Teske}, Johanna},
        title = "{The MIRI Exoplanets Orbiting White dwarfs (MEOW) Survey: Mid-infrared Excess Reveals a Giant Planet Candidate around a Nearby White Dwarf}",
      journal = {\apjl},
         year = 2024,
        month = sep,
       volume = {973},
       number = {1},
          eid = {L11},
        pages = {L11},
          doi = {10.3847/2041-8213/ad74ed},
archivePrefix = {arXiv},
       eprint = {2408.16813},
 primaryClass = {astro-ph.EP},
       adsurl = {https://ui.adsabs.harvard.edu/abs/2024ApJ...973L..11L}
}

@ARTICLE{obrien2024,
       author = {{O'Brien}, Mairi W. and {Tremblay}, P. -E. and {Klein}, B.~L. and {Koester}, D. and {Melis}, C. and {B{\'e}dard}, A. and {Cukanovaite}, E. and {Cunningham}, T. and {Doyle}, A.~E. and {G{\"a}nsicke}, B.~T. and {Gentile Fusillo}, N.~P. and {Hollands}, M.~A. and {McCleery}, J. and {Pelisoli}, I. and {Toonen}, S. and {Weinberger}, A.~J. and {Zuckerman}, B.},
        title = "{The 40 pc sample of white dwarfs from Gaia}",
      journal = {\mnras},
         year = 2024,
        month = jan,
       volume = {527},
       number = {3},
        pages = {8687-8705},
          doi = {10.1093/mnras/stad3773},
archivePrefix = {arXiv},
       eprint = {2312.02735},
 primaryClass = {astro-ph.SR},
       adsurl = {https://ui.adsabs.harvard.edu/abs/2024MNRAS.527.8687O}
}

@ARTICLE{blouin2019,
       author = {{Blouin}, S. and {Dufour}, P. and {Thibeault}, C. and {Allard}, N.~F.},
        title = "{A New Generation of Cool White Dwarf Atmosphere Models. IV. Revisiting the Spectral Evolution of Cool White Dwarfs}",
      journal = {\apj},
         year = 2019,
        month = jun,
       volume = {878},
       number = {1},
          eid = {63},
        pages = {63},
          doi = {10.3847/1538-4357/ab1f82},
archivePrefix = {arXiv},
       eprint = {1905.02174},
 primaryClass = {astro-ph.SR},
       adsurl = {https://ui.adsabs.harvard.edu/abs/2019ApJ...878...63B}
}

@ARTICLE{bedard2017,
       author = {{B{\'e}dard}, A. and {Bergeron}, P. and {Fontaine}, G.},
        title = "{Measurements of Physical Parameters of White Dwarfs: A Test of the Mass-Radius Relation}",
      journal = {\apj},
         year = 2017,
        month = oct,
       volume = {848},
       number = {1},
          eid = {11},
        pages = {11},
          doi = {10.3847/1538-4357/aa8bb6},
archivePrefix = {arXiv},
       eprint = {1709.02324},
 primaryClass = {astro-ph.SR},
       adsurl = {https://ui.adsabs.harvard.edu/abs/2017ApJ...848...11B}
}

@ARTICLE{rodrigo2024,
       author = {{Rodrigo}, Carlos and {Cruz}, Patricia and {Aguilar}, John F. and {Aller}, Alba and {Solano}, Enrique and {G{\'a}lvez-Ortiz}, Maria Cruz and {Jim{\'e}nez-Esteban}, Francisco and {Mas-Buitrago}, Pedro and {Bayo}, Amelia and {Cort{\'e}s-Contreras}, Miriam and {Murillo-Ojeda}, Raquel and {Bonoli}, Silvia and {Cenarro}, Javier and {Dupke}, Renato and {L{\'o}pez-Sanjuan}, Carlos and {Mar{\'\i}n-Franch}, Antonio and {de Oliveira}, Claudia Mendes and {Moles}, Mariano and {Taylor}, Keith and {Varela}, Jes{\'u}s and {Rami{\'o}}, H{\'e}ctor V{\'a}zquez},
        title = "{Photometric segregation of dwarf and giant FGK stars using the SVO Filter Profile Service and photometric tools}",
      journal = {\aap},
         year = 2024,
        month = sep,
       volume = {689},
          eid = {A93},
        pages = {A93},
          doi = {10.1051/0004-6361/202449998},
archivePrefix = {arXiv},
       eprint = {2406.03310},
 primaryClass = {astro-ph.SR},
       adsurl = {https://ui.adsabs.harvard.edu/abs/2024A&A...689A..93R}
}

@software{lmfit,
  author       = {Newville, Matthew and
                  Otten, Renee and
                  Nelson, Andrew and
                  Stensitzki, Till and
                  Ingargiola, Antonino and
                  Allan, Daniel and
                  Fox, Austin and
                  Carter, Faustin and
                  Rawlik, Michal},
  title        = {LMFIT: Non-Linear Least-Squares Minimization and
                   Curve-Fitting for Python
                  },
  month        = mar,
  year         = 2025,
  publisher    = {Zenodo},
  version      = {1.3.3},
  doi          = {10.5281/zenodo.15014437},
  url          = {https://doi.org/10.5281/zenodo.15014437},
  swhid        = {swh:1:dir:6dbda1361832412880d315cbf608ba498c177d82
                   ;origin=https://doi.org/10.5281/zenodo.598352;visi
                   t=swh:1:snp:0d236d4d8ff7f7248600297b145d047734607d
                   14;anchor=swh:1:rel:5f2a70f7c390d9b78cb661b349b044
                   753cf7e89e;path=lmfit-lmfit-py-f97ddf7
                  },
}

@ARTICLE{mustill2014,
       author = {{Mustill}, Alexander J. and {Veras}, Dimitri and {Villaver}, Eva},
        title = "{Long-term evolution of three-planet systems to the post-main sequence and beyond}",
      journal = {\mnras},
         year = 2014,
        month = jan,
       volume = {437},
       number = {2},
        pages = {1404-1419},
          doi = {10.1093/mnras/stt1973},
archivePrefix = {arXiv},
       eprint = {1310.3168},
 primaryClass = {astro-ph.EP},
       adsurl = {https://ui.adsabs.harvard.edu/abs/2014MNRAS.437.1404M}
}

@article{veras2013,
    author = {Veras, Dimitri and Mustill, Alexander J. and Bonsor, Amy and Wyatt, Mark C.},
    title = {Simulations of two-planet systems through all phases of stellar evolution: implications for the instability boundary and white dwarf pollution},
    journal = {Monthly Notices of the Royal Astronomical Society},
    volume = {431},
    number = {2},
    pages = {1686-1708},
    year = {2013},
    month = {03},
    issn = {0035-8711},
    doi = {10.1093/mnras/stt289},
    url = {https://doi.org/10.1093/mnras/stt289},
    eprint = {https://academic.oup.com/mnras/article-pdf/431/2/1686/4278156/stt289.pdf},
}

@ARTICLE{veras2016,
       author = {{Veras}, Dimitri},
        title = "{Post-main-sequence planetary system evolution}",
      journal = {Royal Society Open Science},
         year = 2016,
        month = feb,
       volume = {3},
          eid = {150571},
        pages = {150571},
          doi = {10.1098/rsos.150571},
archivePrefix = {arXiv},
       eprint = {1601.05419},
 primaryClass = {astro-ph.EP},
       adsurl = {https://ui.adsabs.harvard.edu/abs/2016RSOS....350571V}
}

@ARTICLE{limbach2025,
       author = {{Limbach}, Mary Anne and {Vanderburg}, Andrew and {MacDonald}, Ryan J. and {Stevenson}, Kevin B. and {Jenkins}, Sydney and {Blouin}, Simon and {Rauscher}, Emily and {Bowens-Rubin}, Rachel and {Gallo}, Elena and {Mang}, James and {Morley}, Caroline V. and {Sing}, David K. and {O'Connor}, Christopher and {Venner}, Alexander and {Xu}, Siyi},
        title = "{Thermal Emission and Confirmation of the Frigid White Dwarf Exoplanet WD 1856+534 b}",
      journal = {\apjl},
         year = 2025,
        month = may,
       volume = {984},
       number = {1},
          eid = {L28},
        pages = {L28},
          doi = {10.3847/2041-8213/adc9ad},
archivePrefix = {arXiv},
       eprint = {2504.16982},
 primaryClass = {astro-ph.EP},
       adsurl = {https://ui.adsabs.harvard.edu/abs/2025ApJ...984L..28L}
}

@ARTICLE{fontaine2001,
       author = {{Fontaine}, G. and {Brassard}, P. and {Bergeron}, P.},
        title = "{The Potential of White Dwarf Cosmochronology}",
      journal = {\pasp},
         year = 2001,
        month = apr,
       volume = {113},
       number = {782},
        pages = {409-435},
          doi = {10.1086/319535},
       adsurl = {https://ui.adsabs.harvard.edu/abs/2001PASP..113..409F}
}

@ARTICLE{zhang2024,
       author = {{Zhang}, Keming and {Zang}, Weicheng and {El-Badry}, Kareem and {Lu}, Jessica R. and {Bloom}, Joshua S. and {Agol}, Eric and {Gaudi}, B. Scott and {Konopacky}, Quinn and {LeBaron}, Natalie and {Mao}, Shude and {Terry}, Sean},
        title = "{An Earth-mass planet and a brown dwarf in orbit around a white dwarf}",
      journal = {Nature Astronomy},
         year = 2024,
        month = dec,
       volume = {8},
        pages = {1575-1582},
          doi = {10.1038/s41550-024-02375-9},
archivePrefix = {arXiv},
       eprint = {2409.02157},
 primaryClass = {astro-ph.EP},
       adsurl = {https://ui.adsabs.harvard.edu/abs/2024NatAs...8.1575Z}
}

@INPROCEEDINGS{vogt1994,
       author = {{Vogt}, S.~S. and {Allen}, S.~L. and {Bigelow}, B.~C. and {Bresee}, L. and {Brown}, B. and {Cantrall}, T. and {Conrad}, A. and {Couture}, M. and {Delaney}, C. and {Epps}, H.~W. and {Hilyard}, D. and {Hilyard}, D.~F. and {Horn}, E. and {Jern}, N. and {Kanto}, D. and {Keane}, M.~J. and {Kibrick}, R.~I. and {Lewis}, J.~W. and {Osborne}, J. and {Pardeilhan}, G.~H. and {Pfister}, T. and {Ricketts}, T. and {Robinson}, L.~B. and {Stover}, R.~J. and {Tucker}, D. and {Ward}, J. and {Wei}, M.~Z.},
        title = "{HIRES: the high-resolution echelle spectrometer on the Keck 10-m Telescope}",
    booktitle = {Instrumentation in Astronomy VIII},
         year = 1994,
       editor = {{Crawford}, David L. and {Craine}, Eric R.},
       series = {Society of Photo-Optical Instrumentation Engineers (SPIE) Conference Series},
       volume = {2198},
        month = jun,
        pages = {362},
          doi = {10.1117/12.176725},
       adsurl = {https://ui.adsabs.harvard.edu/abs/1994SPIE.2198..362V}
}

@ARTICLE{winston2020,
       author = {{Winston}, Elaine and {Hora}, Joseph L. and {Tolls}, Volker},
        title = "{A Census of Star Formation in the Outer Galaxy. II. The GLIMPSE360 Field}",
      journal = {\aj},
         year = 2020,
        month = aug,
       volume = {160},
       number = {2},
          eid = {68},
        pages = {68},
          doi = {10.3847/1538-3881/ab99c8},
archivePrefix = {arXiv},
       eprint = {2006.03080},
 primaryClass = {astro-ph.GA},
       adsurl = {https://ui.adsabs.harvard.edu/abs/2020AJ....160...68W}
}

@INPROCEEDINGS{debes2025,
       author = {{Debes}, John and {Poulsen}, Sabrina and {Messier}, Ashley and {Cracraft}, Misty and {Mullally}, Susan and {Thibault}, Katherine and {Albert}, Loic and {Dufour}, Patrick and {Barclay}, Thomas and {Hermes}, JJ and {Kilic}, Mukremin and {Lafreniere}, David and {Mullally}, Fergal and {Reach}, William and {Quintana}, Elisa},
        title = "{Metal Polluted White Dwarfs with 21-micron Excesses: Dust or Planets?}",
    booktitle = {American Astronomical Society Meeting Abstracts \#245},
         year = 2025,
       series = {American Astronomical Society Meeting Abstracts},
       volume = {245},
        month = jan,
          eid = {439.06},
        pages = {439.06},
       adsurl = {https://ui.adsabs.harvard.edu/abs/2025AAS...24543906D}
}

@ARTICLE{chambers2016,
       author = {{Chambers}, K.~C. and {Magnier}, E.~A. and {Metcalfe}, N. and {Flewelling}, H.~A. and {Huber}, M.~E. and {Waters}, C.~Z. and {Denneau}, L. and {Draper}, P.~W. and {Farrow}, D. and {Finkbeiner}, D.~P. and {Holmberg}, C. and {Koppenhoefer}, J. and {Price}, P.~A. and {Rest}, A. and {Saglia}, R.~P. and {Schlafly}, E.~F. and {Smartt}, S.~J. and {Sweeney}, W. and {Wainscoat}, R.~J. and {Burgett}, W.~S. and {Chastel}, S. and {Grav}, T. and {Heasley}, J.~N. and {Hodapp}, K.~W. and {Jedicke}, R. and {Kaiser}, N. and {Kudritzki}, R. -P. and {Luppino}, G.~A. and {Lupton}, R.~H. and {Monet}, D.~G. and {Morgan}, J.~S. and {Onaka}, P.~M. and {Shiao}, B. and {Stubbs}, C.~W. and {Tonry}, J.~L. and {White}, R. and {Ba{\~n}ados}, E. and {Bell}, E.~F. and {Bender}, R. and {Bernard}, E.~J. and {Boegner}, M. and {Boffi}, F. and {Botticella}, M.~T. and {Calamida}, A. and {Casertano}, S. and {Chen}, W. -P. and {Chen}, X. and {Cole}, S. and {Deacon}, N. and {Frenk}, C. and {Fitzsimmons}, A. and {Gezari}, S. and {Gibbs}, V. and {Goessl}, C. and {Goggia}, T. and {Gourgue}, R. and {Goldman}, B. and {Grant}, P. and {Grebel}, E.~K. and {Hambly}, N.~C. and {Hasinger}, G. and {Heavens}, A.~F. and {Heckman}, T.~M. and {Henderson}, R. and {Henning}, T. and {Holman}, M. and {Hopp}, U. and {Ip}, W. -H. and {Isani}, S. and {Jackson}, M. and {Keyes}, C.~D. and {Koekemoer}, A.~M. and {Kotak}, R. and {Le}, D. and {Liska}, D. and {Long}, K.~S. and {Lucey}, J.~R. and {Liu}, M. and {Martin}, N.~F. and {Masci}, G. and {McLean}, B. and {Mindel}, E. and {Misra}, P. and {Morganson}, E. and {Murphy}, D.~N.~A. and {Obaika}, A. and {Narayan}, G. and {Nieto-Santisteban}, M.~A. and {Norberg}, P. and {Peacock}, J.~A. and {Pier}, E.~A. and {Postman}, M. and {Primak}, N. and {Rae}, C. and {Rai}, A. and {Riess}, A. and {Riffeser}, A. and {Rix}, H.~W. and {R{\"o}ser}, S. and {Russel}, R. and {Rutz}, L. and {Schilbach}, E. and {Schultz}, A.~S.~B. and {Scolnic}, D. and {Strolger}, L. and {Szalay}, A. and {Seitz}, S. and {Small}, E. and {Smith}, K.~W. and {Soderblom}, D.~R. and {Taylor}, P. and {Thomson}, R. and {Taylor}, A.~N. and {Thakar}, A.~R. and {Thiel}, J. and {Thilker}, D. and {Unger}, D. and {Urata}, Y. and {Valenti}, J. and {Wagner}, J. and {Walder}, T. and {Walter}, F. and {Watters}, S.~P. and {Werner}, S. and {Wood-Vasey}, W.~M. and {Wyse}, R.},
        title = "{The Pan-STARRS1 Surveys}",
      journal = {arXiv e-prints},
         year = 2016,
        month = dec,
          eid = {arXiv:1612.05560},
        pages = {arXiv:1612.05560},
          doi = {10.48550/arXiv.1612.05560},
archivePrefix = {arXiv},
       eprint = {1612.05560},
 primaryClass = {astro-ph.IM},
       adsurl = {https://ui.adsabs.harvard.edu/abs/2016arXiv161205560C}
}

@BOOK{cutri2003,
       author = {{Cutri}, R.~M. and {Skrutskie}, M.~F. and {van Dyk}, S. and {Beichman}, C.~A. and {Carpenter}, J.~M. and {Chester}, T. and {Cambresy}, L. and {Evans}, T. and {Fowler}, J. and {Gizis}, J. and {Howard}, E. and {Huchra}, J. and {Jarrett}, T. and {Kopan}, E.~L. and {Kirkpatrick}, J.~D. and {Light}, R.~M. and {Marsh}, K.~A. and {McCallon}, H. and {Schneider}, S. and {Stiening}, R. and {Sykes}, M. and {Weinberg}, M. and {Wheaton}, W.~A. and {Wheelock}, S. and {Zacarias}, N.},
        title = "{2MASS All Sky Catalog of point sources.}",
         year = 2003,
       adsurl = {https://ui.adsabs.harvard.edu/abs/2003tmc..book.....C}
}

@INPROCEEDINGS{dufour2017,
       author = {{Dufour}, P. and {Blouin}, S. and {Coutu}, S. and {Fortin-Archambault}, M. and {Thibeault}, C. and {Bergeron}, P. and {Fontaine}, G.},
        title = "{The Montreal White Dwarf Database: A Tool for the Community}",
    booktitle = {20th European White Dwarf Workshop},
         year = 2017,
       editor = {{Tremblay}, P. -E. and {Gaensicke}, B. and {Marsh}, T.},
       series = {Astronomical Society of the Pacific Conference Series},
       volume = {509},
        month = mar,
        pages = {3},
          doi = {10.48550/arXiv.1610.00986},
archivePrefix = {arXiv},
       eprint = {1610.00986},
 primaryClass = {astro-ph.SR},
       adsurl = {https://ui.adsabs.harvard.edu/abs/2017ASPC..509....3D}
}

@ARTICLE{eisenhardt2020,
       author = {{Eisenhardt}, Peter R.~M. and {Marocco}, Federico and {Fowler}, John W. and {Meisner}, Aaron M. and {Kirkpatrick}, J. Davy and {Garcia}, Nelson and {Jarrett}, Thomas H. and {Koontz}, Renata and {Marchese}, Elijah J. and {Stanford}, S. Adam and {Caselden}, Dan and {Cushing}, Michael C. and {Cutri}, Roc M. and {Faherty}, Jacqueline K. and {Gelino}, Christopher R. and {Gonzalez}, Anthony H. and {Mainzer}, Amanda and {Mobasher}, Bahram and {Schlegel}, David J. and {Stern}, Daniel and {Teplitz}, Harry I. and {Wright}, Edward L.},
        title = "{The CatWISE Preliminary Catalog: Motions from WISE and NEOWISE Data}",
      journal = {\apjs},
         year = 2020,
        month = apr,
       volume = {247},
       number = {2},
          eid = {69},
        pages = {69},
          doi = {10.3847/1538-4365/ab7f2a},
archivePrefix = {arXiv},
       eprint = {1908.08902},
 primaryClass = {astro-ph.IM},
       adsurl = {https://ui.adsabs.harvard.edu/abs/2020ApJS..247...69E}
}

@ARTICLE{Gaia2023,
       author = {{Gaia Collaboration} and {Vallenari}, A. and {Brown}, A.~G.~A. and {Prusti}, T. and {de Bruijne}, J.~H.~J. and {Arenou}, F. and {Babusiaux}, C. and {Biermann}, M. and {Creevey}, O.~L. and {Ducourant}, C. and {Evans}, D.~W. and {Eyer}, L. and {Guerra}, R. and {Hutton}, A. and {Jordi}, C. and {Klioner}, S.~A. and {Lammers}, U.~L. and {Lindegren}, L. and {Luri}, X. and {Mignard}, F. and {Panem}, C. and {Pourbaix}, D. and {Randich}, S. and {Sartoretti}, P. and {Soubiran}, C. and {Tanga}, P. and {Walton}, N.~A. and {Bailer-Jones}, C.~A.~L. and {Bastian}, U. and {Drimmel}, R. and {Jansen}, F. and {Katz}, D. and {Lattanzi}, M.~G. and {van Leeuwen}, F. and {Bakker}, J. and {Cacciari}, C. and {Casta{\~n}eda}, J. and {De Angeli}, F. and {Fabricius}, C. and {Fouesneau}, M. and {Fr{\'e}mat}, Y. and {Galluccio}, L. and {Guerrier}, A. and {Heiter}, U. and {Masana}, E. and {Messineo}, R. and {Mowlavi}, N. and {Nicolas}, C. and {Nienartowicz}, K. and {Pailler}, F. and {Panuzzo}, P. and {Riclet}, F. and {Roux}, W. and {Seabroke}, G.~M. and {Sordo}, R. and {Th{\'e}venin}, F. and {Gracia-Abril}, G. and {Portell}, J. and {Teyssier}, D. and {Altmann}, M. and {Andrae}, R. and {Audard}, M. and {Bellas-Velidis}, I. and {Benson}, K. and {Berthier}, J. and {Blomme}, R. and {Burgess}, P.~W. and {Busonero}, D. and {Busso}, G. and {C{\'a}novas}, H. and {Carry}, B. and {Cellino}, A. and {Cheek}, N. and {Clementini}, G. and {Damerdji}, Y. and {Davidson}, M. and {de Teodoro}, P. and {Nu{\~n}ez Campos}, M. and {Delchambre}, L. and {Dell'Oro}, A. and {Esquej}, P. and {Fern{\'a}ndez-Hern{\'a}ndez}, J. and {Fraile}, E. and {Garabato}, D. and {Garc{\'\i}a-Lario}, P. and {Gosset}, E. and {Haigron}, R. and {Halbwachs}, J. -L. and {Hambly}, N.~C. and {Harrison}, D.~L. and {Hern{\'a}ndez}, J. and {Hestroffer}, D. and {Hodgkin}, S.~T. and {Holl}, B. and {Jan{\ss}en}, K. and {Jevardat de Fombelle}, G. and {Jordan}, S. and {Krone-Martins}, A. and {Lanzafame}, A.~C. and {L{\"o}ffler}, W. and {Marchal}, O. and {Marrese}, P.~M. and {Moitinho}, A. and {Muinonen}, K. and {Osborne}, P. and {Pancino}, E. and {Pauwels}, T. and {Recio-Blanco}, A. and {Reyl{\'e}}, C. and {Riello}, M. and {Rimoldini}, L. and {Roegiers}, T. and {Rybizki}, J. and {Sarro}, L.~M. and {Siopis}, C. and {Smith}, M. and {Sozzetti}, A. and {Utrilla}, E. and {van Leeuwen}, M. and {Abbas}, U. and {{\'A}brah{\'a}m}, P. and {Abreu Aramburu}, A. and {Aerts}, C. and {Aguado}, J.~J. and {Ajaj}, M. and {Aldea-Montero}, F. and {Altavilla}, G. and {{\'A}lvarez}, M.~A. and {Alves}, J. and {Anders}, F. and {Anderson}, R.~I. and {Anglada Varela}, E. and {Antoja}, T. and {Baines}, D. and {Baker}, S.~G. and {Balaguer-N{\'u}{\~n}ez}, L. and {Balbinot}, E. and {Balog}, Z. and {Barache}, C. and {Barbato}, D. and {Barros}, M. and {Barstow}, M.~A. and {Bartolom{\'e}}, S. and {Bassilana}, J. -L. and {Bauchet}, N. and {Becciani}, U. and {Bellazzini}, M. and {Berihuete}, A. and {Bernet}, M. and {Bertone}, S. and {Bianchi}, L. and {Binnenfeld}, A. and {Blanco-Cuaresma}, S. and {Blazere}, A. and {Boch}, T. and {Bombrun}, A. and {Bossini}, D. and {Bouquillon}, S. and {Bragaglia}, A. and {Bramante}, L. and {Breedt}, E. and {Bressan}, A. and {Brouillet}, N. and {Brugaletta}, E. and {Bucciarelli}, B. and {Burlacu}, A. and {Butkevich}, A.~G. and {Buzzi}, R. and {Caffau}, E. and {Cancelliere}, R. and {Cantat-Gaudin}, T. and {Carballo}, R. and {Carlucci}, T. and {Carnerero}, M.~I. and {Carrasco}, J.~M. and {Casamiquela}, L. and {Castellani}, M. and {Castro-Ginard}, A. and {Chaoul}, L. and {Charlot}, P. and {Chemin}, L. and {Chiaramida}, V. and {Chiavassa}, A. and {Chornay}, N. and {Comoretto}, G. and {Contursi}, G. and {Cooper}, W.~J. and {Cornez}, T. and {Cowell}, S. and {Crifo}, F. and {Cropper}, M. and {Crosta}, M. and {Crowley}, C. and {Dafonte}, C. and {Dapergolas}, A. and {David}, M. and {David}, P. and {de Laverny}, P. and {De Luise}, F. and {De March}, R.},
        title = "{Gaia Data Release 3. Summary of the content and survey properties}",
      journal = {\aap},
         year = 2023,
        month = jun,
       volume = {674},
          eid = {A1},
        pages = {A1},
          doi = {10.1051/0004-6361/202243940},
archivePrefix = {arXiv},
       eprint = {2208.00211},
 primaryClass = {astro-ph.GA},
       adsurl = {https://ui.adsabs.harvard.edu/abs/2023A&A...674A...1G}
}

@ARTICLE{bergeron2019,
       author = {{Bergeron}, P. and {Dufour}, P. and {Fontaine}, G. and {Coutu}, S. and {Blouin}, S. and {Genest-Beaulieu}, C. and {B{\'e}dard}, A. and {Rolland}, B.},
        title = "{On the Measurement of Fundamental Parameters of White Dwarfs in the Gaia Era}",
      journal = {\apj},
         year = 2019,
        month = may,
       volume = {876},
       number = {1},
          eid = {67},
        pages = {67},
          doi = {10.3847/1538-4357/ab153a},
archivePrefix = {arXiv},
       eprint = {1904.02022},
 primaryClass = {astro-ph.SR},
       adsurl = {https://ui.adsabs.harvard.edu/abs/2019ApJ...876...67B}
}

@BOOK{press1986,
       author = {{Press}, William H. and {Flannery}, Brian P. and {Teukolsky}, Saul A.},
        title = "{Numerical recipes. The art of scientific computing}",
         year = 1986,
       adsurl = {https://ui.adsabs.harvard.edu/abs/1986nras.book.....P}
}

@ARTICLE{reach2005,
       author = {{Reach}, William T. and {Kuchner}, Marc J. and {von Hippel}, Ted and {Burrows}, Adam and {Mullally}, Fergal and {Kilic}, Mukremin and {Winget}, D.~E.},
        title = "{The Dust Cloud around the White Dwarf G29-38}",
      journal = {\apjl},
         year = 2005,
        month = dec,
       volume = {635},
       number = {2},
        pages = {L161-L164},
          doi = {10.1086/499561},
archivePrefix = {arXiv},
       eprint = {astro-ph/0511358},
 primaryClass = {astro-ph},
       adsurl = {https://ui.adsabs.harvard.edu/abs/2005ApJ...635L.161R}
}

@ARTICLE{becklin2005,
       author = {{Becklin}, E.~E. and {Farihi}, J. and {Jura}, M. and {Song}, Inseok and {Weinberger}, A.~J. and {Zuckerman}, B.},
        title = "{A Dusty Disk around GD 362, a White Dwarf with a Uniquely High Photospheric Metal Abundance}",
      journal = {\apjl},
         year = 2005,
        month = oct,
       volume = {632},
       number = {2},
        pages = {L119-L122},
          doi = {10.1086/497826},
archivePrefix = {arXiv},
       eprint = {astro-ph/0509193},
 primaryClass = {astro-ph},
       adsurl = {https://ui.adsabs.harvard.edu/abs/2005ApJ...632L.119B}
}

@software{bushouse2023,
  author       = {Bushouse, Howard and
                  Eisenhamer, Jonathan and
                  Dencheva, Nadia and
                  Davies, James and
                  Greenfield, Perry and
                  Morrison, Jane and
                  Hodge, Phil and
                  Simon, Bernie and
                  Grumm, David and
                  Droettboom, Michael and
                  Slavich, Edward and
                  Sosey, Megan and
                  Pauly, Tyler and
                  Miller, Todd and
                  Jedrzejewski, Robert and
                  Hack, Warren and
                  Davis, David and
                  Crawford, Steven and
                  Law, David and
                  Gordon, Karl and
                  Regan, Michael and
                  Cara, Mihai and
                  MacDonald, Ken and
                  Bradley, Larry and
                  Shanahan, Clare and
                  Jamieson, William and
                  Teodoro, Mairan and
                  Williams, Thomas},
  title        = {JWST Calibration Pipeline},
  month        = jun,
  year         = 2023,
  publisher    = {Zenodo},
  version      = {1.11.1},
  doi          = {10.5281/zenodo.8099867},
  url          = {https://doi.org/10.5281/zenodo.8099867},
}

@ARTICLE{mullally2025,
       author = {{Mullally}, Fergal and {Mullally}, Susan E. and {Cracraft}, Misty and {Bianco}, Samantha N. and {Albert}, Loic and {Debes}, John and {Hermes}, J.~J. and {Kilic}, Mukremin and {Reach}, William T.},
        title = "{Follow-up Observations of Candidate White Dwarf Planets with MIRI}",
      journal = {arXiv e-prints},
         year = 2025,
        month = dec,
          eid = {arXiv:2512.08191},
        pages = {arXiv:2512.08191},
          doi = {10.48550/arXiv.2512.08191},
archivePrefix = {arXiv},
       eprint = {2512.08191},
 primaryClass = {astro-ph.SR},
       adsurl = {https://ui.adsabs.harvard.edu/abs/2025arXiv251208191M}
}

@ARTICLE{loic2025,
       author = {{Albert}, Lo{\"\i}c and {Poulsen}, Sabrina R. and {Le Bourdais}, {\'E}rika and {Debes}, John H. and {Boucher}, Samuel and {Kilic}, Mukremin and {Reach}, William and {Mullally}, Susan E. and {Cracraft}, Misty and {Mullally}, Fergal and {De Furio}, Matthew and {Hermes}, J.~J. and {Kenyon}, Scott J. and {Melis}, Carl and {Redfield}, Seth and {Wyatt}, M.~C. and {Dufour}, Patrick and {Golimowski}, David A. and {Messier}, Ashley and {Farihi}, Jay},
        title = "{The MIRI Excesses around Degenerates (MEAD) Survey I: A candidate cold brown dwarf in orbit around the nearby white dwarf 2MASS J09424023-4637176}",
      journal = {arXiv e-prints},
         year = 2025,
        month = oct,
          eid = {arXiv:2510.12601},
        pages = {arXiv:2510.12601},
          doi = {10.48550/arXiv.2510.12601},
archivePrefix = {arXiv},
       eprint = {2510.12601},
 primaryClass = {astro-ph.SR},
       adsurl = {https://ui.adsabs.harvard.edu/abs/2025arXiv251012601A}
}

@ARTICLE{reach2025,
       author = {{Reach}, William T. and {Kilic}, Mukremin and {Lisse}, Carey M. and {Debes}, John H. and {von Hippel}, Ted and {Azartash-Namin}, Bianca and {Lo{\"\i}c Albert} and {Mullally}, Susan E. and {Mullally}, Fergal and {Cracraft}, Misty and {Bernice}, Madison and {Erickson}, Selin L.},
        title = "{Composition of Planetary Debris Around the White Dwarf GD 362}",
      journal = {\apj},
         year = 2025,
        month = dec,
       volume = {994},
       number = {2},
          eid = {195},
        pages = {195},
          doi = {10.3847/1538-4357/ae11a9},
archivePrefix = {arXiv},
       eprint = {2510.07595},
 primaryClass = {astro-ph.GA},
       adsurl = {https://ui.adsabs.harvard.edu/abs/2025ApJ...994..195R}
}

@ARTICLE{farihi2025,
       author = {{Farihi}, J. and {Su}, K.~Y.~L. and {Melis}, C. and {Kenyon}, S.~J. and {Swan}, A. and {Redfield}, S. and {Wyatt}, M.~C. and {Debes}, J.~H.},
        title = "{Subtle and Spectacular: Diverse White Dwarf Debris Disks Revealed by JWST}",
      journal = {\apjl},
         year = 2025,
        month = mar,
       volume = {981},
       number = {1},
          eid = {L5},
        pages = {L5},
          doi = {10.3847/2041-8213/adae88},
archivePrefix = {arXiv},
       eprint = {2501.18338},
 primaryClass = {astro-ph.EP},
       adsurl = {https://ui.adsabs.harvard.edu/abs/2025ApJ...981L...5F}
}

@ARTICLE{sanderson2022,
       author = {{Sanderson}, Hannah and {Bonsor}, Amy and {Mustill}, Alexander},
        title = "{Can Gaia find planets around white dwarfs?}",
      journal = {\mnras},
         year = 2022,
        month = dec,
       volume = {517},
       number = {4},
        pages = {5835-5852},
          doi = {10.1093/mnras/stac2867},
archivePrefix = {arXiv},
       eprint = {2206.02505},
 primaryClass = {astro-ph.EP},
       adsurl = {https://ui.adsabs.harvard.edu/abs/2022MNRAS.517.5835S}
}

@ARTICLE{pourbaix2022,
       author = {{Pourbaix}, D. and {Arenou}, F. and {Gavras}, P. and {Gosset}, {\'E}. and {Halbwachs}, J.-L. and {Siopis}, C. and {Sozzetti}, A. and {Bauchet}, N. and {Damerdji}, Y. and {Delchambre}, L. and {Delisle}, J.-B. and {Giacobbe}, P. and {Holl}, B. and {Jorissen}, A. and {Lattanzi}, M.~G. and {Leclerc}, N. and {Morel}, T. and {Sadowski}, G. and {Sahlmann}, J. and {Segransan}, D.},
        title = "{Gaia DR3 documentation Chapter 7: Non-single stars}",
 howpublished = {Gaia DR3 documentation, European Space Agency; Gaia Data Processing and Analysis Consortium. Online at <A href=``https://gea.esac.esa.int/archive/documentation/GDR3/index.html''>https://gea.esac.esa.int/archive/documentation/GDR3/index.html</A>, id. 7},
         year = 2022,
        month = jun,
          eid = {7},
        pages = {7},
       adsurl = {https://ui.adsabs.harvard.edu/abs/2022gdr3.reptE...7P}
}
\bibliographystyle{aasjournal}



\end{document}